\documentclass[lettersize,journal]{IEEEtran}
\usepackage{amsmath,amsfonts}
\usepackage{algorithmic}
\usepackage{algorithm}
\usepackage{array}
\usepackage[table,dvipsnames]{xcolor}
\usepackage[caption=false,font=normalsize,labelfont=sf,textfont=sf]{subfig}
\usepackage{multirow}
\usepackage{textcomp}
\usepackage{stfloats}
\usepackage{makecell}
\usepackage{url}
\usepackage{booktabs}
\usepackage{cite}
\usepackage{verbatim}
\usepackage{graphicx}
\usepackage{threeparttable}
\usepackage{pifont}

\newcommand\xy{\textcolor{black}}

\begin{document}

\title{SpikeAid: A Streaming Spiking Hearing Aid Framework for Joint Noise Reduction and Wide Dynamic Range Compression}

\author{
Xiang~Hao,
Xinyi~Chen,
Hanwen~Liu,
Richard~Hau~Yue~So,
Kay~Chen~Tan,
and Jibin~Wu%
\thanks{
Xiang Hao, Xinyi Chen, Jibin Wu, and Kay Chen Tan
are with The Hong Kong Polytechnic University,
Hong Kong SAR, China.
}%
\thanks{
Hanwen Liu is with the University of Electronic
Science and Technology of China,
Chengdu, China.
}%
\thanks{
Richard Hau Yue So is with
The Hong Kong University of Science and Technology,
Hong Kong SAR, China.
}%
\thanks{
Corresponding author: Xinyi Chen
(e-mail: xinyi-97.chen@connect.polyu.hk).
}%
}



\maketitle

\begin{abstract}
Modern hearing aids combine noise reduction (NR) with wide dynamic range compression (WDRC) to suppress background interference while compensating for the reduced auditory dynamic range associated with hearing loss. Conventional systems typically process NR and WDRC as separate stages, although the output of NR directly affects the signal levels and envelopes that determine subsequent compression. Recent end-to-end neural approaches can jointly optimize these processes, but their dense computation remains challenging for always-on hearing aids.
In this work, we propose SpikeAid, a causal end-to-end spiking neural network (SNN) framework for joint NR and WDRC. SpikeAid consists of an NR-Net that constructs complementary speech and noise representations and a WDRC-Net that estimates the corresponding branch gains, with both components jointly optimized according to the final hearing aid objective. To capture the multiscale temporal structures required by joint NR and WDRC in a computationally efficient manner, we further introduce the Delayed Resonate-and-fire Neuron (DRN), which incorporates bio-inspired heterogeneous dendritic delays and resonating dynamics.
Experimental results on the hearing aid task show that SpikeAid improves key objective metrics over cascaded processing and outperforms existing spiking speech-processing baselines.
Evaluation on the public Intel Neuromorphic Deep Noise Suppression (N-DNS) Challenge further demonstrates state-of-the-art neuromorphic speech-denoising performance with DRN. Finally, we designed an FPGA-based neuromorphic hardware system for efficient deployment of SpikeAid, where it achieves over two orders of magnitude higher energy efficiency than conventional CPU/GPU implementations, demonstrating its potential for energy-efficient hearing aid applications.
\end{abstract}

\begin{IEEEkeywords}
Hearing Aids, Spiking Neural Networks, Noise Reduction, Wide Dynamic Range
Compression, Neuromorphic Computing.
\end{IEEEkeywords}

\section{Introduction}
\IEEEPARstart{H}{earing} loss presents a growing global public-health challenge. The World Health Organization estimates that by 2050, nearly 2.5 billion people will experience some degree of hearing loss, with more than 700 million requiring hearing rehabilitation~\cite{who-world-report-hearing-2021}. For many individuals with permanent sensorineural hearing loss, damaged cochlear sensory cells cannot be restored through routine clinical treatment. Hearing aids therefore remain the principal non-surgical means of improving speech audibility and everyday communication~\cite{cunningham-2017-hearing-loss-adults}. This creates a fundamental need for hearing aid processing methods that improve speech intelligibility while supporting efficient always-on streaming operation.

\begin{figure}[!t]
\centering
\includegraphics[width=\columnwidth]{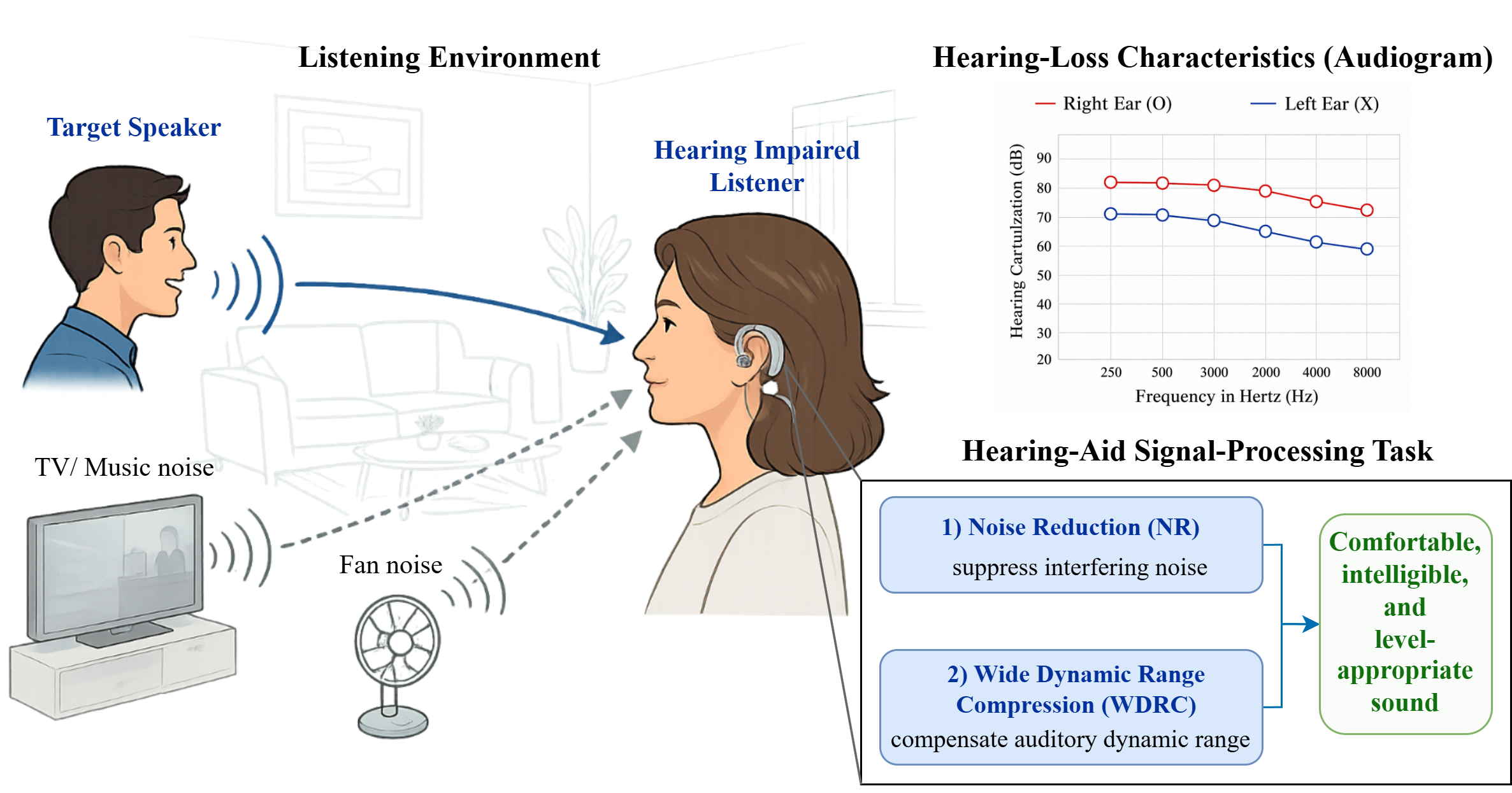}
\caption{\textbf{Illustration of the hearing aid signal-processing task.} The hearing aid receives target speech mixed with environmental interference in a noisy listening environment. An audiogram depicts the listener's frequency-dependent hearing loss. Within the hearing aid, NR suppresses background interference, whereas WDRC applies level-dependent amplification to accommodate the reduced auditory dynamic range associated with hearing loss. Together, these processes aim to produce an output that is intelligible, comfortable, and at an appropriate level.}
\label{fig:hearing-aid-task}
\end{figure}

To support effective speech communication in everyday acoustic environments, hearing aids must address two different problems: background interference and the reduced auditory dynamic range caused by sensorineural hearing loss, as illustrated in Fig.~\ref{fig:hearing-aid-task}. Noise reduction (NR) suppresses interfering sounds to improve speech perception and listening comfort, but it does not compensate for the listener's hearing loss. This compensation is commonly determined from an audiogram, which describes the listener's frequency-dependent hearing loss. A hearing aid fitting prescription then uses the audiogram to define a prescribed wide dynamic range compression (WDRC) profile, specifying the desired amplification at different frequencies and input levels. In particular, WDRC provides greater gain to low-level sounds and progressively reduces the gain as the input level increases, thereby mapping the wide range of environmental sound levels into the listener's narrower residual auditory dynamic range~\cite{plomp-1978-limited-benefit-hearing-aids,kowalewski-2020-perceptual-snr-aware-drc}. Conventional hearing aids therefore typically arrange NR and WDRC in a cascade, with NR followed by WDRC~\cite{ngo-2012-combined-nr-drc,yun-2024-nr-wdrc-output-snr,zhang-2025-neural-wdrc}.

Although this cascaded design is simple and modular, NR and WDRC do not operate independently. NR changes which speech and noise components remain in the signal, as well as their levels and temporal envelopes. Because the subsequent WDRC gain is determined from these signal characteristics, the output of NR directly affects the compression behavior. WDRC then further modifies both the preserved speech and any residual interference. As a result, low-energy speech components suppressed by NR cannot be recovered by WDRC, while residual noise may be amplified together with the desired speech. WDRC can also further reshape the output signal-to-noise ratio (SNR) and speech-envelope characteristics. SNR-aware and source-aware compression strategies partially account for these interactions by adapting compression according to speech or noise information~\cite{may-2018-snr-aware-drc,yun-2024-nr-wdrc-output-snr}, but generally retain separately designed processing stages. More recently, neural approaches such as Neural-WDRC have demonstrated the potential of jointly optimizing noise suppression and compression within a unified model~\cite{zhang-2025-neural-wdrc}.


Joint optimization addresses the interaction between NR and WDRC, but it does not by itself resolve the deployment constraints of practical hearing aids.
To support continuous use in a compact wearable device, hearing aid signal-processing algorithms must operate in real time under stringent latency, memory, and energy constraints.
Existing neural joint-processing approaches rely on dense, frame-wise artificial neural network (ANN) computation, where multiply-accumulate (MAC) operations are continuously performed regardless of the instantaneous acoustic activity. This leads to a challenging quality--latency--power trade-off for real-time deployment~\cite{drgas-2023-low-latency-dnn-se}. These constraints motivate efficient alternatives that are better aligned with low-power hearing aid hardware.

Spiking neural networks (SNNs) provide a promising computing paradigm inspired by the computational principles of biological neural systems. Instead of using continuous activations, SNNs transmit and process information through discrete spike events \cite{maass1997networks}. This spike-based representation naturally supports sparse and asynchronous computation, allowing neurons to remain inactive unless informative events arrive or spikes are generated. As a result, SNNs can reduce unnecessary computation and memory access, making them attractive for low-power streaming applications \cite{ma2017darwin,balaji2019mapping,rathi2023exploring,chen2023hybrid,PLIF,10.1609/aaai.v40i3.37174}. Beyond efficiency, SNNs also possess rich neuronal dynamics: their internal states evolve over time in response to incoming spikes, enabling temporal information to be integrated and retained across successive time steps \cite{ma2025spiking,10.24963/ijcai.2025/544,adLIF,dhsnn,PSN,chen2026temporal}. These properties make SNNs particularly suitable for hearing aid processing, which requires continuous tracking of speech and noise envelope dynamics.
Recent SNN-based speech enhancement studies have shown that spiking architectures can achieve competitive denoising performance while targeting low-power or low-latency operation. For example, Spiking-FullSubNet uses full-band and sub-band spiking modeling for ultralow-power neuromorphic speech enhancement \cite{hao-2025-spiking-fullsubnet}, and Dual-Path Spiking Neural Network (DPSNN) develops a streaming time-domain SNN for low-latency speech enhancement \cite{sun-2024-dpsnn}.  These studies show that sparse spiking computation can achieve competitive speech enhancement performance.

Despite recent progress in neural hearing aid processing and SNN-based speech enhancement, several challenges remain before SNNs can support joint NR-WDRC processing in practical hearing aids. First, existing SNN speech enhancement systems are primarily designed for standalone denoising, with the objective of reconstructing clean speech. To the best of our knowledge, no end-to-end SNN framework has yet been developed to produce a hearing aid output jointly determined by noise reduction and a prescribed WDRC profile. Second, joint NR and WDRC require modeling temporal structures across multiple timescales. NR relies on both short-term spectro-temporal structures and longer contextual information about speech and interference, whereas WDRC must track level trajectories to provide responsive yet stable gain control. Capturing these heterogeneous temporal structures while retaining efficient training and causal streaming inference remains challenging for existing spiking models. Third, the practical hardware efficiency of such a neuromorphic hearing aid system remains insufficiently demonstrated. It remains unclear whether such a system can be deployed to achieve low-power and low-latency streaming inference.

To address these challenges, we propose SpikeAid, a causal end-to-end SNN framework for joint NR and WDRC. SpikeAid jointly optimizes NR-Net and WDRC-Net according to the final hearing aid objective. To model the temporal structures required by joint NR and WDRC, we introduce \xy{a spiking neuron model tailored for streaming speech enhancement}, termed Delayed Resonate-and-fire Neuron (DRN), which combines bio-inspired heterogeneous dendritic delays with resonating dynamics. We further implement SpikeAid on FPGA-based neuromorphic hardware to evaluate its practical deployment and streaming-inference efficiency.

The main contributions of this work are summarized as follows: 

\begin{itemize}
\item We propose SpikeAid, the first causal end-to-end SNN framework that jointly integrates NR and WDRC into a unified streaming model, directly transforming noisy speech into hearing-compensated output. 
\item We introduce DRN, a bio-inspired spiking neuron model underlying both NR-Net and WDRC-Net. It effectively models temporal structures across multiple timescales while supporting efficient parallel training and causal streaming inference.
\item Experiments on the hearing aid task show that SpikeAid achieves the best overall performance among the evaluated SNN baselines, with a favorable trade-off among speech quality, hearing-aid-oriented perceptual performance, and computational efficiency. Evaluation on the public Intel N-DNS Challenge further demonstrates the effectiveness of DRN, achieving state-of-the-art speech-denoising performance.
\item We implement SpikeAid on FPGA-based neuromorphic hardware, achieving over 100$\times$ and 800$\times$ higher energy efficiency than CPU and GPU implementations, respectively, demonstrating its practical potential for hearing aid devices that require operations under stringent energy constraints.
\end{itemize}

\section{Related Work}

\subsection{Joint NR and WDRC for Hearing Aids}

Early studies coordinated NR and WDRC through shared speech-presence or signal-to-noise-ratio information. Ngo et al. coupled multichannel Wiener-filter NR with WDRC to mitigate the output-SNR degradation of conventional cascaded processing \cite{ngo-2012-combined-nr-drc}. SNR-aware WDRC subsequently employed different compression speeds in speech- and noise-dominated regions, providing perceptual advantages over fixed-speed compression \cite{may-2018-snr-aware-drc,kowalewski-2020-perceptual-snr-aware-drc}. Later investigations further showed that applying NR before WDRC does not necessarily prevent compression-induced degradation of the output SNR, highlighting the need to consider the two stages jointly \cite{yun-2024-nr-wdrc-output-snr,overby-2025-preference-nr-wdrc}. These approaches improve the coordination between NR and WDRC, but generally rely on separately designed modules or predefined compression policies. They therefore do not provide an end-to-end framework whose output is jointly determined by NR and a prescribed WDRC profile.

Neural hearing-loss compensation instead learns listener-dependent processing from audiograms, fitting prescriptions, or differentiable auditory models. Representative systems include differentiable hearing aid fitting \cite{tu-2021-dhasp}, auditory-model-based individualized compensation \cite{drakopoulos-2023-individualized-hlc}, and interpretable neural dynamic processors that approximate multiband compression \cite{drgas-2024-dynamic-hlc,leer-2025-dnn-hlc}. More closely related to the present work, several methods jointly perform speech enhancement and hearing-loss compensation using cascaded neural modules \cite{tu-2021-two-stage}, audiogram-conditioned enhancement networks \cite{cheng-2023-ftcrn-hlc,ni-2026-afn-hearnet}, or source-dependent neural compression \cite{zhang-2025-neural-wdrc,zhang-2025-audiogram-nr-wdrc}. Neural-WDRC, in particular, estimates speech and noise components and applies source-specific compression, demonstrating perceptual advantages over conventional and SNR-aware WDRC in nonstationary noise \cite{zhang-2025-neural-wdrc}. However, these systems primarily depend on continuously active dense ANN inference, while the effectiveness and efficiency of the complete joint NR-WDRC chain have rarely been validated at the hardware level. 

\subsection{SNN-based Speech Enhancement}

SNNs provide an event-driven neural computing paradigm in which information is represented and transmitted through discrete spike events \cite{maass1997networks}. With surrogate-gradient learning, SNNs can be trained for complex sequence-processing tasks while retaining sparse spike-based activity \cite{neftci-2019-surrogate-gradient,li2021differentiable, Xu_2023_CVPR, pingtnnls2025,ma2026spatiotemporal}. This property has motivated the use of SNNs for low-power audio and speech processing, especially in scenarios where real-time operation and energy efficiency are important. The Intel Neuromorphic Deep Noise Suppression (N-DNS) Challenge established a common benchmark for neuromorphic speech enhancement, including an algorithmic track and a Loihi~2 hardware track evaluated in terms of output quality, latency, energy, and resource consumption \cite{timcheck_intel_2023}. Riahi and Plourde trained a U-Net SNN with surrogate gradients and reported performance above the challenge baseline and competitive with an analogous ANN \cite{riahi-2023-unet-snn}. Spiking Structured State Space Model (Spiking-S4) combined spiking computation with a structured state-space model to capture long-range dependencies, achieving competitive enhancement with fewer parameters and FLOPs than representative ANN systems \cite{chen2024spikingS4}. Spiking-FullSubNet combines full-band and sub-band modeling with frequency partitioning and adaptive spiking dynamics, and won the algorithmic track of the Intel N-DNS Challenge \cite{hao-2025-spiking-fullsubnet}. Other studies explored spiking-transformer bottlenecks \cite{alohali-2024-spiking-transformer} and hybrid ANN-to-SNN fine-tuning \cite{abuhajar-2025-hybrid-snn-se}. Collectively, these methods address long-context modeling, spectral representation, low-latency streaming, and computational efficiency. 
However, they are formulated primarily as standalone denoisers that reconstruct clean speech or estimate enhancement masks. They neither produce a listener-dependent output governed by a prescribed WDRC profile nor optimize how compression reshapes enhanced speech and residual interference. 
In addition, their temporal mechanisms are designed mainly for speech denoising, rather than for jointly modeling the multiscale acoustic dynamics required by NR and the level-dependent gain trajectories introduced by WDRC.
Moreover, although the N-DNS hardware track has established the feasibility of neuromorphic SNN-based speech enhancement, a complete causal spiking system for joint NR and prescribed WDRC has not yet been validated on neuromorphic hardware. 

\subsection{Dendritic-inspired Spiking Neuron Models}

\xy{Recent studies have explored biologically inspired spiking neurons with richer internal dynamics to improve temporal processing beyond conventional Leaky Integrate-and-Fire (LIF) neurons \cite{lif}. One line of research focuses on resonate-and-fire dynamics \cite{pmlr-v235-higuchi24a,zhang2026dendritic}, where intrinsic oscillatory behaviors are introduced into spiking neurons to capture temporal structures across different timescales. Another line of research investigates dendritic-inspired mechanisms to enrich temporal representation through structural heterogeneity. The Dendritic Heterogeneity Spiking Neural Network (DH-SNN) \cite{zheng2024temporal} introduces multiple dendritic branches with different temporal constants, enabling heterogeneous temporal integration across branches. The Parallel Multi-compartment Spiking Neuron (PMSN) \cite{11604167} models dendritic dynamics based on cable theory. The Multi-branch Multi-compartment Dendritic-inspired spiking neuron (MMDend) \cite{wang2025mmdend} further explores dendritic structures with multiple branches and compartments to enhance temporal representation through richer neuronal dynamics.}

\xy{Although these approaches significantly improve the temporal modeling capability of spiking neurons, streaming hearing aid processing imposes more specific challenges. Acoustic signals contain both rapidly varying transient structures, such as speech onsets, consonant transitions, and noise bursts. Existing neuron models do not explicitly model heterogeneous temporal delays caused by different dendritic pathways, limiting their ability to construct diverse short-term temporal receptive fields for fine-grained acoustic structures. Meanwhile, several studies have investigated fast Fourier transform (FFT)-based parallel training strategies for these dendritic-inspired spiking neurons \cite{11604167,wang2025mmdend,zhang2026dendritic}. These FFT-based approaches accelerate temporal computation by approximating recurrent dynamics as convolution in the frequency domain. However, this approximation can lead to a mismatch between the neuronal dynamics used during parallel training and those realized during streaming inference. Such a mismatch is particularly concerning in streaming hearing-aid applications, where models must process continuous input sequences far longer than the segments used for training.}

\section{Proposed SpikeAid Framework}
\label{sec:spikeaid_method}

\begin{figure*}[!t]
\centering
\includegraphics[width=\textwidth]{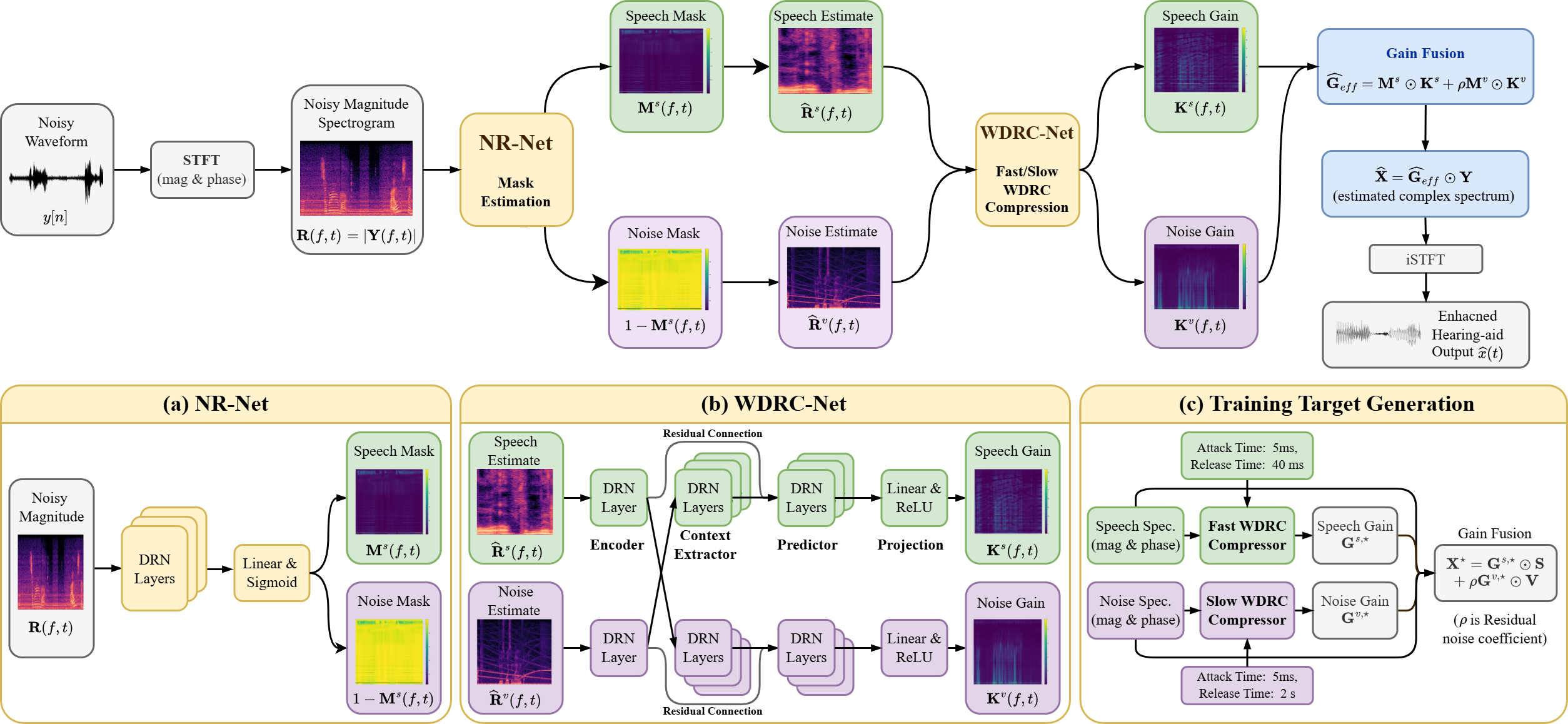}
\caption{\textbf{Overview of SpikeAid.} NR-Net estimates complementary speech and noise masks from the noisy magnitude spectrum and uses them to construct the corresponding branch representations. WDRC-Net uses information from both branches to estimate nonnegative branch gains. The gain fusion stage combines the masks and branch gains into the effective gain, which is applied to the noisy complex spectrum before inverse STFT reconstruction.}
\label{fig:spikeaid-overview}
\end{figure*}

This section introduces SpikeAid from the task formulation to its implementation. We first define the task and summarize the end-to-end processing pipeline. We then introduce the DRN, which serves as the basic temporal processing unit in SpikeAid. Next, we describe the two neural components, NR-Net and WDRC-Net, and explain how their outputs are combined to compute the effective gain and reconstruct the output waveform. \xy{We subsequently describe how the training target is generated using a source-independent WDRC procedure, where speech and noise are processed separately by their respective WDRC operators before recombination, and define the end-to-end loss function.} Finally, we present the causal streaming implementation of the complete framework.

\subsection{Problem Formulation and Framework Overview}
\label{sec:problem_overview}

We consider a monaural mixture $y[n]=s[n]+v[n]$, where $s[n]$ and $v[n]$ denote the target speech and additive noise, respectively. Let $L$ denote the number of time-domain samples and $n\in\{0,\ldots,L-1\}$ denote the sample index. Let $\mathbf{Y}=\operatorname{STFT}(y)$, $\mathbf{S}=\operatorname{STFT}(s)$, and $\mathbf{V}=\operatorname{STFT}(v)$ denote their complex short-time Fourier transform (STFT) spectra. By linearity of the STFT, $\mathbf{Y}=\mathbf{S}+\mathbf{V}$, where $\mathbf{Y},\mathbf{S},\mathbf{V}\in\mathbb{C}^{F\times T}$. Here, $F$ and $T$ are the numbers of frequency bins and time frames, respectively; $f\in\{0,\ldots,F-1\}$ and $t\in\{0,\ldots,T-1\}$ index frequency and time; and $\odot$ denotes element-wise multiplication.

Unlike conventional denoising, the desired hearing aid output must reflect both NR and a prescribed WDRC profile. Following the source-independent WDRC formulation in~\cite{zhang-2025-neural-wdrc}, in which speech and noise are compressed separately before recombination, we define the training target $\mathbf{X}^{\star} \in\mathbb{C}^{F\times T}$ as
\begin{equation}
\mathbf{X}^{\star} = \mathbf{G}^{s,\star}\odot\mathbf{S} + \rho \mathbf{G}^{v,\star}\odot\mathbf{V}, \qquad \rho=10^{-\eta_{\mathrm{NR}}/20}, 
\label{eq:source_independent_target}
\end{equation}
where $\mathbf{G}^{s,\star},\mathbf{G}^{v,\star} \in\mathbb{R}_{+}^{F\times T}$ are the gains for speech and noise, respectively, as specified by the prescribed WDRC profile. Each element of these prescribed WDRC gains specifies the multiplicative gain applied at a particular frequency bin and time frame: low-level components generally receive greater gain, whereas high-level components receive less gain, thereby compressing the acoustic dynamic range. The parameter $\eta_{\mathrm{NR}}$ specifies the residual-noise attenuation in decibels, and $\rho$ is its linear-amplitude equivalent. The superscript $\star$ identifies variables associated with the training target. The prescribed WDRC gains are generated offline from the clean speech and noise spectra, as detailed in Section~\ref{sec:target_objective}.

In this work, we propose the SpikeAid framework shown in Fig.~\ref{fig:spikeaid-overview}. SpikeAid contains two neural components, NR-Net and WDRC-Net. The framework takes the noisy magnitude spectrum $\mathbf{R}=|\mathbf{Y}|\in\mathbb{R}_{+}^{F\times T}$ as input. NR-Net estimates complementary speech and noise masks, $\mathbf{M}^{s}$ and $\mathbf{M}^{v}$, and uses them to construct the corresponding branch representations, $\widetilde{\mathbf{R}}^{s}$ and $\widetilde{\mathbf{R}}^{v}$. WDRC-Net then processes these two branch representations, exchanges temporal context between the branches, and estimates the nonnegative branch gains $\mathbf{K}^{s}$ and $\mathbf{K}^{v}$. After the two neural components, a gain fusion stage combines the masks and branch gains to obtain the effective gain. The reconstruction stage applies the effective gain to the noisy complex spectrum $\mathbf{Y}$ and uses an inverse STFT to reconstruct the output waveform. The complete system is trained end to end to approximate the training target defined in Eq.~\eqref{eq:source_independent_target}.

Having outlined the overall architecture and signal flow of SpikeAid, we next introduce the spiking neuron model underlying both NR-Net and WDRC-Net. 

\subsection{Delayed Resonate-and-fire Neuron (DRN)}
\label{sec:drn}



\begin{figure*}[!t]
\centering
\includegraphics[width=\textwidth]{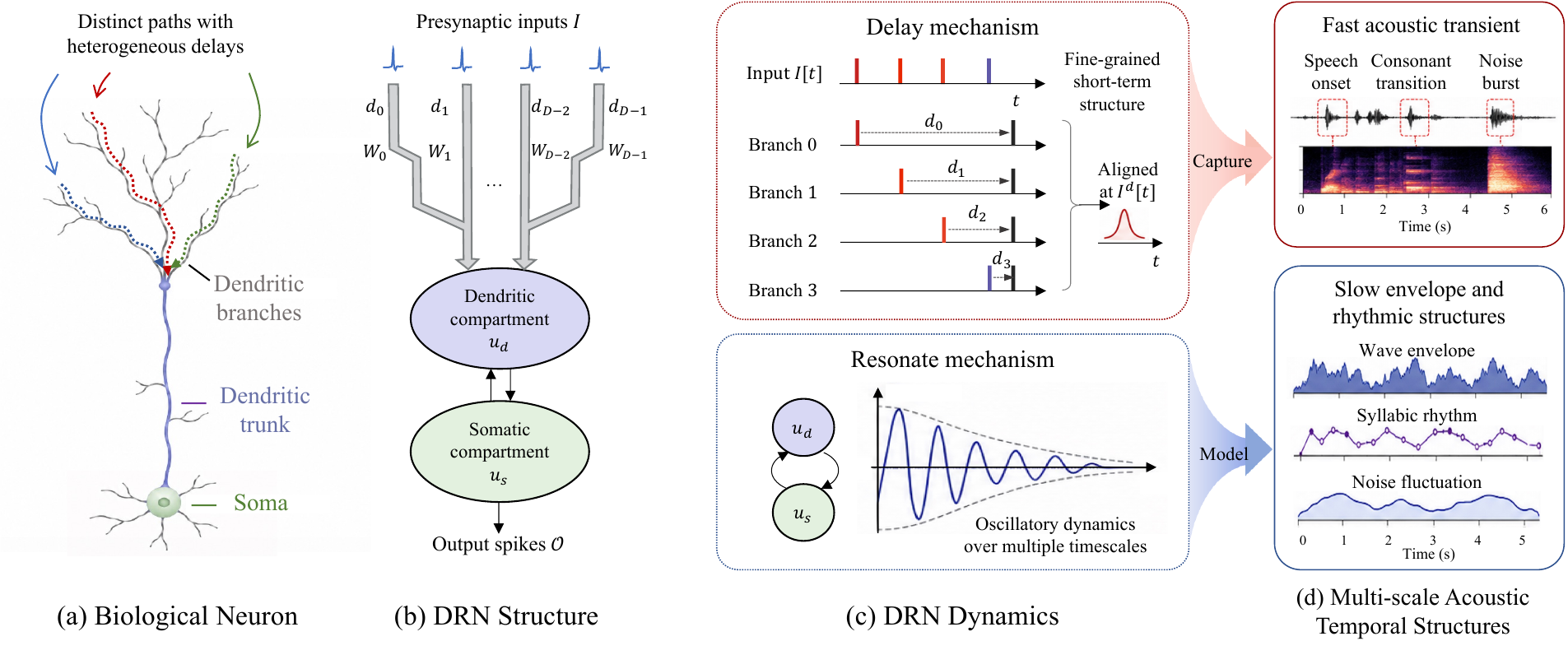}
\caption{\textbf{Illustration of the Delayed Resonate-and-fire Neuron (DRN).} Heterogeneously delayed inputs are aggregated by the dendritic compartment, while coupled dendritic–somatic dynamics produce damped resonating states. Together, these mechanisms capture local acoustic transients and longer-term envelope dynamics.}
\label{fig:drn}
\end{figure*}

\subsubsection{Design Motivation and Neuronal Dynamics}

\xy{Streaming hearing aid processing requires models to simultaneously capture temporal structures across multiple timescales. Short-term temporal variations, such as speech onsets, consonant transitions, and noise bursts, contain fine-grained acoustic information, whereas long-term temporal evolution, including speech envelopes and WDRC gain trajectories, requires persistent temporal memory. Motivated by these temporal requirements, we introduce the DRN, a spiking neuron model tailored for streaming speech enhancement, which integrates heterogeneous dendritic delays with resonating dendrite--soma dynamics to achieve multi-timescale temporal modeling.}

As illustrated in Fig.~\ref{fig:drn}, biological neurons contain rich internal structures, including dendritic branches, dendritic trunk, and the soma, that support temporal processing beyond simple membrane integration and threshold-based firing.
When a presynaptic input arrives, it is first transmitted through dendritic branches before reaching the dendritic trunk. Since different branches correspond to different distal-to-proximal transmission paths, the incoming signals may arrive with heterogeneous delays, allowing the neuron to detect and retain fine-grained temporal information within a local time window.
The dendritic compartment subsequently interacts with the soma through bidirectional coupling. Dendritic activity drives the somatic response, while the somatic state feeds back to modulate the dendritic compartment. This interaction can produce damped subthreshold oscillations, enabling the neuron to model temporal structures over
extended timescales.

Inspired by these biological observations, we propose the DRN model, which is designed to incorporate two key components: First, the delay mechanism enables the neuron to model intricate transient temporal structures in speech and noise signals, such as speech onsets, consonant transitions, and noise bursts. Second, the resonate mechanism introduces oscillatory subthreshold dynamics, allowing the neuron to model temporal structures over extended timescales, such as envelope modulations, syllabic rhythms, and slowly varying noise fluctuations. By combining these two mechanisms, DRN can capture both fast acoustic transients and longer-range envelope dynamics, providing a suitable spiking unit for joint NR-WDRC processing.

Given a presynaptic input current $I(t)$, the delayed dendritic input is obtained by aggregating $D$ dendritic branches:
\begin{align}
I^{d}(t) = \sum_{k=0}^{D-1} W_k I(t-d_k),
\label{eq:drf_delay}
\end{align}
where $d_k$ and $W_k$ denote the delay and learnable weight of the $k$th dendritic branch pathway, respectively.
This delayed aggregation allows the dendritic compartment to receive temporally dispersed presynaptic information rather than only the instantaneous input.

Let $u_{d}(t)$ and $u_{s}(t)$ denote the dendritic and somatic membrane potentials, respectively. Their coupled subthreshold dynamics are defined as
\begin{equation}
\frac{d u_d(t)}{dt} = \alpha u_d(t) - \beta u_s(t) + I^{d}(t),
\label{eq:drn_dendritic_dynamic}
\end{equation}
\begin{equation}
\frac{d u_s(t)}{dt} = \alpha u_s(t) + \beta u_d(t) ,\hspace{10mm}
\label{eq:drn_somatic_dynamic}
\end{equation}
where parameter $\alpha<0$ controls the decay rate of both membrane potentials and $\beta>0$ controls the coupling strength between the dendritic and somatic compartments. Eq. \eqref{eq:drn_dendritic_dynamic} describes the integration of delayed inputs and somatic feedback in the dendritic compartment, whereas Eq.~\eqref{eq:drn_somatic_dynamic} describes the response of the soma to dendritic activity. Their bidirectional coupling gives rise to damped resonating dynamics.

This property can be seen more directly by defining the complex state
\begin{equation}
z(t) = u_d(t) + \boldsymbol{i} u_s(t),
\label{eq:drf_complex_state}
\end{equation}
where $\boldsymbol{i}$ is the imaginary unit. Eqs. \eqref{eq:drn_dendritic_dynamic}--\eqref{eq:drn_somatic_dynamic} can then be written compactly as
\begin{equation}
\frac{d z(t)}{dt} = (\alpha + \boldsymbol{i}\beta)z(t) + I^{d}(t).
\label{eq:drn_complex_dynamic}
\end{equation}
This form is analogous to the classical resonate-and-fire neuron\cite{izhikevich2001resonate}. 
Specifically, $\alpha$ acts as a damping factor that exponentially decays the oscillation amplitude, while $\beta$ determines the intrinsic oscillatory frequency. 

The somatic potential is then integrated over time for spike generation:
\begin{align}
\frac{d m(t)}{dt} = u_s(t),
\label{eq:drf_membrane}
\end{align}
where $m(t)$ denotes the firing membrane potential. An output spike
is generated when it reaches the firing threshold:
\begin{align}
\mathcal{O}(t) = H\big(m(t)-\theta\big),
\label{eq:drf_spike}
\end{align}
where $H(\cdot)$ is the Heaviside step function and $\theta$ is the firing threshold. After firing, the membrane potential is reset as
\begin{equation}
    m(t) \leftarrow m(t) - r(t)  \theta,   
\label{eq:drf_reset}
\end{equation}
where $r(t)=\lfloor m(t)/\theta \rfloor$ denotes the integer reset factor.

\subsubsection{Stable Discretization and Multiscale Temporal Modeling}




The continuous-time formulation above defines the temporal dynamics of DRN. To support practical training and inference, we next derive a stable discrete-time realization and characterize how it enables temporal modeling over multiple timescales.

For the $j$th DRN, let $\lambda_j=\alpha_j+\mathrm{i}\beta_j$ denote its complex coefficient. Applying zero-order-hold discretization to Eq.~\eqref{eq:drn_complex_dynamic} with step size $\Delta_j>0$ gives 
\begin{equation} 
z_j[t] = A_j z_j[t-1]+B_j I_j^{d}[t], 
\label{eq:drn_discrete_dynamic} 
\end{equation} 
where 
\begin{equation} 
A_j = \exp(\Delta_j\lambda_j), \qquad B_j = \frac{\exp(\Delta_j\lambda_j)-1}{\lambda_j}. \label{eq:drn_zoh} 
\end{equation} 

To keep the resonant dynamics stable during optimization, we parameterize the real part of $\alpha_j$ using an unconstrained learnable parameter $a_j$ as
\begin{equation} 
\alpha_j=-\exp(a_j). 
\label{eq:drn_stable_parameterization} 
\end{equation} 
This guarantees $|A_j|=\exp(\Delta_j\alpha_j)<1$ for any $\Delta_j>0$ and therefore prevents divergent resonating states.

Beyond stability, DRN captures temporal structures through two complementary mechanisms. First, the delay mechanism provides neuron-specific temporal filtering within a short causal window. The $D$ transmission delays are initialized as
\begin{equation}
    d_k=k,
    \qquad
    k=0,1,\ldots,D-1,
    \label{eq:drn_delay_initialization}
\end{equation}
and the delayed input of the $j$th neuron is therefore
\begin{equation}
    I_j^{d}[t]
    =
    \sum_{k=0}^{D-1}W_{j,k}I[t-k].
    \label{eq:drn_delay_convolution}
\end{equation}
Each neuron has an independent delay kernel $\{W_{j,k}\}_{k=0}^{D-1}$ and can consequently learn a distinct local temporal filter. These filters enable the DRN population to capture fine-grained acoustic structures within a short time window.


Second, the resonate mechanism provides temporal selectivity over extended timescales. Under the stable parameterization in Eq.~\eqref{eq:drn_stable_parameterization}, the state-transition coefficient can be written as
\begin{equation}
    A_j
    =
    \exp\!\left[-\Delta_j\exp(a_j)\right]
    \exp\!\left(\mathrm{i}\Delta_j\beta_j\right),
    \label{eq:drn_transition_interpretation}
\end{equation}
where the term $\Delta_j\exp(a_j)$ controls the decay speed and effective memory length, whereas $\Delta_j\beta_j$ determines the preferred temporal frequency. Heterogeneous resonating parameters across the DRN population therefore provide a diverse set of damped temporal filters across multiple timescales. Together with the delay mechanism, the DRN layer is able to capture temporal structures across multiple timescales.
Further derivations of the stability condition and the frequency-response analysis of the resonating dynamics are provided in Supplementary Materials. 


\subsubsection{Parallel Training and Streaming Inference}

Hearing aid processing requires efficient training on long audio sequences while ultimately being implemented on an always-on streaming device. To this end, DRN is designed to support parallel computation during training and recurrent streaming inference without changing the underlying state transition.

First, the delay operation in Eq.~\eqref{eq:drn_delay_convolution} is a neuron-wise causal time-invariant convolution. The delayed inputs of all neurons and timesteps can therefore be computed in parallel using a grouped one-dimensional convolution.

Second, we adapt a dedicated associative-scan kernel~\cite{blelloch1990prefix,harris2007parallel} to the resonating dynamics in Eq.~\eqref{eq:drn_discrete_dynamic}. Let $u_j[t]=B_jI_j^{d}[t]$ and represent each transition as $(A_j,u_j[t])$. Two consecutive transitions can be composed as
\begin{equation}
(A_2,u_2)\circ(A_1,u_1)
=
(A_2A_1,\;A_2u_1+u_2),
\end{equation}
which is associative. 
Compared with FFT-based parallel methods that approximate dynamics through full-sequence convolution in the frequency domain \cite{11604167,wang2025mmdend,zhang2026dendritic}, the proposed scan-based implementation preserves the recurrent state transition, thereby avoiding approximation errors that lead to inconsistent neuronal dynamics between parallel training and streaming inference. This consistency is particularly important for streaming hearing-aid processing, where inference sequences can be substantially longer than training clips.


We next describe the overall spiking architecture, which consists of two neural components, NR-Net and WDRC-Net, both built upon the proposed DRN temporal unit.

\subsection{SpikeAid Architecture}
\label{sec:spikeaid_architecture}
The following subsections describe the SpikeAid architecture in three parts. We first introduce NR-Net, then present WDRC-Net, and finally describe the gain fusion and reconstruction stages. For notational convenience, the batch dimension is omitted and the spectral tensors are arranged in an $F\times T$ layout. Because all DRN layers are causal, the same mappings can be evaluated frame by frame during streaming inference. Let $\mathcal{Q}=\{s,v\}$ denote the speech and noise branches, and let $\bar{q}$ denote the branch complementary to $q$, i.e., $\bar{s}=v$ and $\bar{v}=s$.

\subsubsection{NR-Net}
\label{sec:nr_net}
NR-Net takes the noisy magnitude spectrum $\mathbf{R}=|\mathbf{Y}|$ as input and estimates complementary speech and noise masks. The masks and the corresponding branch representations are defined as
\begin{equation}
\begin{aligned} 
\mathbf{M}^{s} &= \operatorname{sigmoid} \left[ \mathcal{P}_{\mathrm{NR}} \left( \mathcal{D}_{\mathrm{NR}}(\mathbf{R}) \right) \right], & \mathbf{M}^{v} &= \mathbf{1}-\mathbf{M}^{s}, \\ \widetilde{\mathbf{R}}^{q} &= \mathbf{M}^{q}\odot\mathbf{R}, && q\in\mathcal{Q}, 
\end{aligned}
\label{eq:complementary_masks} 
\end{equation}
where $\mathcal{D}_{\mathrm{NR}}$ is a stack of DRN layers and $\mathcal{P}_{\mathrm{NR}}$ is a linear projection from the hidden dimension to the $F$ STFT bins. The resulting masks satisfy $\mathbf{M}^{q}\in[0,1]^{F\times T}$ and $\mathbf{M}^{s}+\mathbf{M}^{v}=\mathbf{1}$. Consequently, the branch representations satisfy $\widetilde{\mathbf{R}}^{s}+\widetilde{\mathbf{R}}^{v}=\mathbf{R}$. 
The superscript $q$ indicates the intended role of each branch and does not imply an exact decomposition into clean speech and noise.
In particular, $\widetilde{\mathbf{R}}^{s}$ and $\widetilde{\mathbf{R}}^{v}$  are not directly supervised to match $|\mathbf{S}|$ and $|\mathbf{V}|$. Because supervision is applied only to the final output, errors in the intermediate branch representations may be compensated by the downstream gain estimation and gain fusion stages during end-to-end training.

\subsubsection{WDRC-Net}
\label{sec:wdrc_net}

WDRC-Net takes the two branch representations produced by NR-Net as input. For each branch $q\in\mathcal{Q}$, its representation is encoded as binary hidden features. Temporal context extracted from the other branch is then added to these features, and the resulting representation is used to estimate the nonnegative branch gain $\mathbf{K}^{q}$:
\begin{equation}
	\begin{aligned}
		\mathbf{H}^{q} & = \mathcal{E}_{q}\left( \widetilde{\mathbf{R}}^{q}\right),  \\
		\mathbf{U}^{q} & = \mathbf{H}^{q}+ \mathcal{C}_{\bar{q}\rightarrow q}\left( \mathbf{H}^{\bar{q}}\right), \\
		\mathbf{K}^{q} & = \operatorname{ReLU}\left[ \mathcal{P}_{q}\left( \mathcal{D}_{q}(\mathbf{U}^{q}) \right) \right], \qquad q\in\mathcal{Q}.
	\end{aligned}
	\label{eq:wdrc_net_branch_gains}
\end{equation}
Here, $\mathcal{E}_{q}$ is the encoder for branch $q$ and is implemented as a single DRN layer. The context extractor $\mathcal{C}_{\bar{q}\rightarrow q}$ is implemented as a stack of DRN layers and extracts temporal context from the other branch. The gain predictor $\mathcal{D}_{q}$ is implemented as another stack of DRN layers, and $\mathcal{P}_{q}$ linearly projects its output to the $F$ frequency bins.
By using temporal context from the other branch, WDRC-Net estimates each branch gain using information from both branches. The branch gains satisfy $\mathbf{K}^{q}\in\mathbb{R}_{+}^{F\times T}$. Because no intermediate loss directly supervises them against the prescribed WDRC gains $\mathbf{G}^{q,\star}$, $\mathbf{K}^{q}$ should not be interpreted as a direct estimate of $\mathbf{G}^{q,\star}$. Instead, the branch gains are optimized according to their contribution to the final gain used for reconstruction.

\subsubsection{Gain Fusion and Reconstruction Stages}
\label{sec:gain_fusion}

The gain fusion stage combines the complementary masks produced by NR-Net with the branch gains produced by WDRC-Net to obtain a single gain for output reconstruction:
\begin{equation} 
\widehat{\mathbf{G}}_{\mathrm{eff}} = \mathbf{M}^{s}\odot\mathbf{K}^{s} + \rho\, \mathbf{M}^{v}\odot\mathbf{K}^{v}, 
\label{eq:effective_gain}
\end{equation} 
where $\rho$ is the residual-noise attenuation factor defined in Eq.~\eqref{eq:source_independent_target}. We refer to $\widehat{\mathbf{G}}_{\mathrm{eff}}$ as the effective gain.

The output spectrum and waveform are reconstructed as
\begin{equation}
\begin{aligned}
\widehat{\mathbf{X}} &= \widehat{\mathbf{G}}_{\mathrm{eff}}\odot\mathbf{Y}, \\ \widehat{x}[n] &= \operatorname{iSTFT} \left( \widehat{\mathbf{X}} \right). 
\end{aligned}
\label{eq:spikeaid_reconstruction}
\end{equation}
Because the effective gain is real-valued, applying it to $\mathbf{Y}$ preserves the phase of the noisy mixture spectrum. SpikeAid can therefore approximate the training target in Eq.~\eqref{eq:source_independent_target} without explicitly estimating and recombining separate complex speech and noise spectra. Reconstruction requires only one element-wise multiplication and one inverse STFT. The complete architecture is trained end to end using a loss applied to the reconstructed spectrum. The next subsection describes the offline generation of the training target and the loss function used for training.

\subsection{Training Target and Loss Function}
\label{sec:target_objective}

\subsubsection{Training Target}
\label{sec:training_target}
The training target in Eq.~\eqref{eq:source_independent_target} is generated offline from the clean speech and noise spectra, as illustrated in Fig.~\ref{fig:spikeaid-overview}~(c).
Following the source-independent WDRC formulation in~\cite{zhang-2025-neural-wdrc}, WDRC is applied separately to speech and noise before the two processed spectra are recombined. Here, ``source-independent'' refers to the separate processing of the two sources; the corresponding WDRC operators may still use different temporal parameters.

For training target construction, we use $q\in\mathcal{Q}$ to index the clean speech and noise spectra, with $\mathbf{Q}^{s}=\mathbf{S}$ and $\mathbf{Q}^{v}=\mathbf{V}$. The prescribed WDRC gains are expressed as
\begin{equation}
    \mathbf{G}^{q,\star}
    =
    \mathcal{W}_{\Theta_q}
    \left(
        \mathbf{Q}^{q}
    \right),
    \qquad
    q\in\mathcal{Q},
    \label{eq:prescribed_wdrc_gains}
\end{equation}
where $\mathcal{W}_{\Theta_q}: \mathbb{C}^{F\times T} \rightarrow \mathbb{R}_{+}^{F\times T}$ is a causal multiband WDRC operator that maps a source spectrum to its time-frequency gains. The parameter set $\Theta_q$ specifies the WDRC configuration for source $q$.

Following established source-independent WDRC designs~\cite{zhang-2025-neural-wdrc,overby2023data_driven}, each operator first estimates calibrated short-time levels within the compression bands, tracks the temporal envelopes using the prescribed attack and release time constants, applies the compression curve, and interpolates the resulting band gains to the STFT frequency bins. The speech and noise WDRC operators share the same compression bands, attack time, compression threshold, and compression ratio, but use different release times. A shorter release time is used for speech to restore the gain more quickly after high-level speech events, whereas a longer release time is used for noise to reduce rapid gain fluctuations in the residual background.

The WDRC configuration is reported in Section~\ref{sec:acoustic_frontend_target}. Details of the level calibration, compression-band construction, envelope recursion, numerical stabilization, and frequency interpolation are provided in Supplementary Materials. Together, the prescribed WDRC gains define the training target in Eq.~\eqref{eq:source_independent_target}.

The WDRC operators are used only to generate the training target offline. Neither the clean source spectra nor the prescribed WDRC gains are available during inference. Moreover, the present work uses a fixed WDRC profile to isolate the joint NR-WDRC learning problem; it does not incorporate listener-specific fitting through insertion-gain targets or audiogram conditioning.

\subsubsection{Loss Function}
\label{sec:loss_function}
The loss function is applied only to the final output.
Given the reconstructed spectrum $\widehat{\mathbf{X}}$ and the training target $\mathbf{X}^{\star}$, the loss function combines complex-domain and magnitude-domain reconstruction errors:
\begin{equation}
    \mathcal{L}
    =
    \frac{
        \mu
        \left\|
            \widehat{\mathbf{X}}-\mathbf{X}^{\star}
        \right\|_{F}^{2}
        +
        (1-\mu)
        \left\|
            \left|\widehat{\mathbf{X}}\right|
            -
            \left|\mathbf{X}^{\star}\right|
        \right\|_{F}^{2}
    }{
        \left\|\mathbf{X}^{\star}\right\|_{F}^{2}
        +
        \epsilon_{E}
    },
    \label{eq:spikeaid_loss}
\end{equation}
where $\mu\in[0,1]$ balances the two reconstruction terms and $\epsilon_{E}>0$ prevents numerical instability for low-energy utterances. The complex-domain term penalizes the real and imaginary reconstruction errors, whereas the magnitude-domain term directly penalizes the output-magnitude error and therefore places additional emphasis on the output-level trajectory. The loss is computed independently for each utterance and then averaged over the minibatch.

Because the loss is applied only after gain fusion and reconstruction, NR-Net and WDRC-Net are jointly optimized according to their contribution to the final hearing aid output. Consequently, the intermediate masks and branch gains are not directly supervised to match the clean source spectra or the prescribed WDRC gains.


\section{Experiments and Results}

\subsection{Experimental Setup}
\label{sec:experimental_setup}

This subsection describes the experimental setup for the joint NR-WDRC hearing aid task. 
Unless otherwise stated, the hearing aid systems use the same data-generation procedure, acoustic frontend, training-target generation, loss function, and optimization settings.
In all result tables, \textbf{bold} and \underline{underlined} values denote the best and second-best results, respectively.

\subsubsection{Data Simulation and Evaluation Sets}

Clean speech and environmental noise are drawn from LibriSpeech~\cite{panayotov-2015-librispeech} and the DNS-Challenge full-band noise corpus~\cite{dubey2022icassp}, respectively. All signals are converted to monaural waveforms sampled at 16~kHz. Training samples are generated online by cropping or zero-padding each clean utterance to 3.072~s and mixing it with a randomly selected noise segment of the same duration.

\xy{The clean-speech level is uniformly sampled from $\{55,60,65,70,75\}$~dB SPL. Following Neural-WDRC~\cite{zhang-2025-neural-wdrc}, we use a digital-to-SPL calibration in which a signal with RMS value 1 corresponds to 95~dB SPL. This calibration provides a consistent mapping between digital signal amplitude and the SPL values used by the level-dependent WDRC procedure.} The input SNR is uniformly sampled from the integers in $\{0,1,\ldots,20\}$~dB. If the maximum absolute amplitude of the clean speech, noise, or mixture exceeds 0.99, all three signals are rescaled by the same factor to prevent clipping without changing the SNR. Each epoch contains 150,000 independently generated mixtures.

We generate fixed validation and evaluation sets offline. Each set contains 120 mixtures spanning input SNRs of $-6$, 0, 6, and 12~dB. The clean utterances used in the two sets are mutually disjoint. Speaker gender was not explicitly balanced or used as a selection criterion during set construction. The validation set is used for model selection and early stopping, whereas the evaluation set is reserved for final evaluation.

\subsubsection{Acoustic Frontend and Training Target}
\label{sec:acoustic_frontend_target}

The acoustic frontend uses a 512-point STFT with a 512-sample Hann window and a 256-sample hop. At 16~kHz, these settings correspond to a 32-ms window, a 16-ms frame shift, and $F=257$ one-sided frequency bins.

The training target is generated using the source-independent WDRC procedure described in Section~\ref{sec:training_target}. Separate speech and noise WDRC operators process frequency bands centered at 125, 250, 500, 1000, 2000, 4000, and 8000~Hz. Both operators use a 5-ms attack time, a 40-dB-SPL compression threshold, and a 2:1 compression ratio. The release times are 40~ms for speech and 2~s for noise. The resulting band gains are interpolated to the STFT frequency bins using natural cubic splines. Before recombination, the compressed noise is attenuated by $\eta_{\mathrm{NR}}=15$~dB, corresponding to $\rho=10^{-15/20}$. This fixed WDRC profile is used to study joint NR-WDRC learning rather than listener-specific fitting.

\subsubsection{Network Configuration}

In our implementation, each DRN layer consists of a linear projection, batch normalization, and DRN neurons applied sequentially. Unless otherwise specified, the number of dendritic delay branches is set to $D=7$.
NR-Net comprises four DRN layers with a hidden dimension of 128, followed by a linear projection and sigmoid activation that estimate the speech mask; the noise mask is its complement. WDRC-Net uses a hidden dimension of 192. Each branch contains a one-layer encoder, a four-layer context extractor, and a four-layer gain predictor whose output is projected to the $F=257$ frequency bins to estimate the branch gains. \xy{Detailed optimization, model-selection, and training-hardware settings are provided in Supplementary Materials.}

\subsubsection{Evaluation Metrics}

Perceptual Evaluation of Speech Quality (PESQ)~\cite{itu-p8622} and Short-Time Objective Intelligibility (STOI)~\cite{taal-2010-stoi} are computed between the system output waveform and the waveform reconstructed from the complex training target. 
\xy{Hearing-Aid Speech Perception Index (HASPI) version 2 and Hearing-Aid Speech Quality Index (HASQI) version 2~\cite{kates-2022-haspi-hasqi} are computed following the evaluation configuration of Neural-WDRC~\cite{zhang-2025-neural-wdrc}. A flat 40-dB HL audiogram is used for the simulated hearing-impaired listener, and an RMS value of 1 is calibrated to 95 dB sound pressure level (SPL). For HASQI v2, National Acoustic Laboratories–Revised (NAL-R) equalization is applied to the clean reference according to the metric definition. The system outputs are evaluated without post-hoc level normalization. The same configuration is used for all systems.} Computational efficiency is reported using the Power Proxy in MOp/s, the power--delay-product (PDP) Proxy in MOp, and the estimated Energy Cost in $\mu\mathrm{J}$. Their definitions and operation-cost assumptions follow the Intel N-DNS Challenge methodology~\cite{timcheck_intel_2023}.

\begin{table*}[t]
\centering
\small
\caption{Effect of joint optimization on the hearing aid task. The compared systems vary the NR processing, WDRC processing, and optimization strategy. }
\label{tab:joint_comparison}
\renewcommand{\arraystretch}{1.2}
\resizebox{\textwidth}{!}{%
\begin{tabular}{lccc c cccc}
\toprule
\textbf{System}
& \textbf{NR Processing}
& \textbf{WDRC Processing}
& \textbf{Optimization}
& \textbf{Params (k)}
& \textbf{PESQ $\uparrow$}
& \textbf{STOI $\uparrow$}
& \textbf{HASPI $\uparrow$}
& \textbf{HASQI $\uparrow$} \\
\midrule

Noisy mixture + fixed WDRC
& None
& Fixed
& -- 
& --
& 1.648
& 0.7799
& 0.3866
& 0.1808 \\

Wiener filtering + fixed WDRC
& Wiener filter
& Fixed
& NR only
& --
& 1.579
& 0.7761
& 0.3667
& 0.1724 \\

NR-Net + fixed WDRC
& NR-Net
& Fixed
& NR only
& 987
& \underline{2.091}
& 0.8103
& \underline{0.4254}
& \textbf{0.1819} \\

SpikeAid (stage-wise)
& NR-Net
& Learnable WDRC-Net
& Stage-wise
& 994
& 2.040
& \underline{0.8278}
& 0.4185
& 0.1792 \\

\rowcolor{gray!12}
SpikeAid
& NR-Net
& Learnable WDRC-Net
& Joint
& 994
& \textbf{2.214}
& \textbf{0.8412}
& \textbf{0.4365}
& \underline{0.1812} \\

\bottomrule
\end{tabular}%
}
\end{table*}

\subsection{Effective Hearing Aid Processing through Joint NR-WDRC Optimization}

First, we compare SpikeAid against a series of NR-WDRC baselines in Table~\ref{tab:joint_comparison}, which differ from SpikeAid in the NR processing, WDRC processing, and optimization strategy. PESQ and STOI measure agreement with the prescribed NR-WDRC target, whereas HASPI and HASQI evaluate hearing-aid-oriented perceptual performance. Among all evaluated systems, SpikeAid achieves the strongest overall hearing aid processing results, obtaining the best PESQ, STOI, and HASPI scores while maintaining competitive HASQI performance.

The first group of baselines follows the conventional cascaded NR-WDRC paradigm, in which the WDRC stage is fixed and only the preceding NR processing is varied. ``NR-Net + fixed WDRC'' substantially outperforms both ``Noisy mixture + fixed WDRC'' and ``Wiener filtering + fixed WDRC'' on most metrics, showing that a stronger learned NR frontend can significantly improve cascaded hearing aid processing.

\xy{Compared with the strong ``NR-Net + fixed WDRC" cascade, SpikeAid achieves higher PESQ, STOI, and HASPI, although its HASQI is slightly lower. This result suggests that the advantage of SpikeAid cannot be attributed solely to stronger standalone denoising, and that coordinating NR and WDRC within a unified learnable framework can improve the final hearing aid output. Notably, ``SpikeAid (stage-wise)" obtains a lower PESQ than ``NR-Net + fixed WDRC." These two systems, however, allocate their learnable capacity differently. The fixed-WDRC baseline concentrates essentially all learnable parameters in NR-Net, whereas stage-wise SpikeAid distributes a comparable total parameter budget across NR-Net and WDRC-Net. Therefore, this comparison does not isolate the effect of making WDRC learnable.}

The effect of joint optimization is more directly isolated by comparing SpikeAid with ``SpikeAid (stage-wise)", which uses the same network architecture and parameter allocation but optimizes NR-Net and WDRC-Net separately. Joint optimization improves all four objective metrics, demonstrating that end-to-end coordination between NR-Net and WDRC-Net is a key contributor to the performance of SpikeAid.

\subsection{Improved Target Matching in Soft-Speech Regions}
\label{sec:soft_speech_analysis}

The results in Table~\ref{tab:joint_comparison} show the overall benefit of joint NR-WDRC optimization, but do not indicate where this benefit arises. We therefore examine soft-speech regions, which are particularly vulnerable in cascaded processing. When NR is optimized independently, low-energy but target-relevant speech components may be attenuated before compression. Once these components are suppressed, the subsequent WDRC stage can only adjust the gain of the remaining signal and cannot reconstruct such lost speech information.

We define soft-speech regions relative to the level range within each utterance. Specifically, the RMS level is computed for each time frame, and frames that are 20--40~dB weaker than the strongest frame in the same utterance are selected as soft-speech frames. We then restrict the analysis to the 0.5--4~kHz frequency range within these frames and denote the resulting time-frequency region by $\mathcal{A}_i$:
\begin{equation}
e_i^{(m)}
=
\frac{
\displaystyle\sum_{(f,t)\in\mathcal{A}_i}
\left|
  \left|\widehat{X}_{i,f,t}^{(m)}\right|
  -
  \left|X_{i,f,t}^{\star}\right|
\right|
}{
\displaystyle\sum_{(f,t)\in\mathcal{A}_i}
\left|X_{i,f,t}^{\star}\right|
+\varepsilon_{\mathrm{a}}
}.
\label{eq:soft_speech_target_error}
\end{equation}

The numerator measures the accumulated magnitude mismatch within the selected soft-speech region, while the denominator normalizes it by the target magnitude. Thus, each utterance contributes one relative error value, with a lower value indicating closer agreement with the desired output.

\begin{figure*}[!t]
\centering
\includegraphics[width=0.85\linewidth]{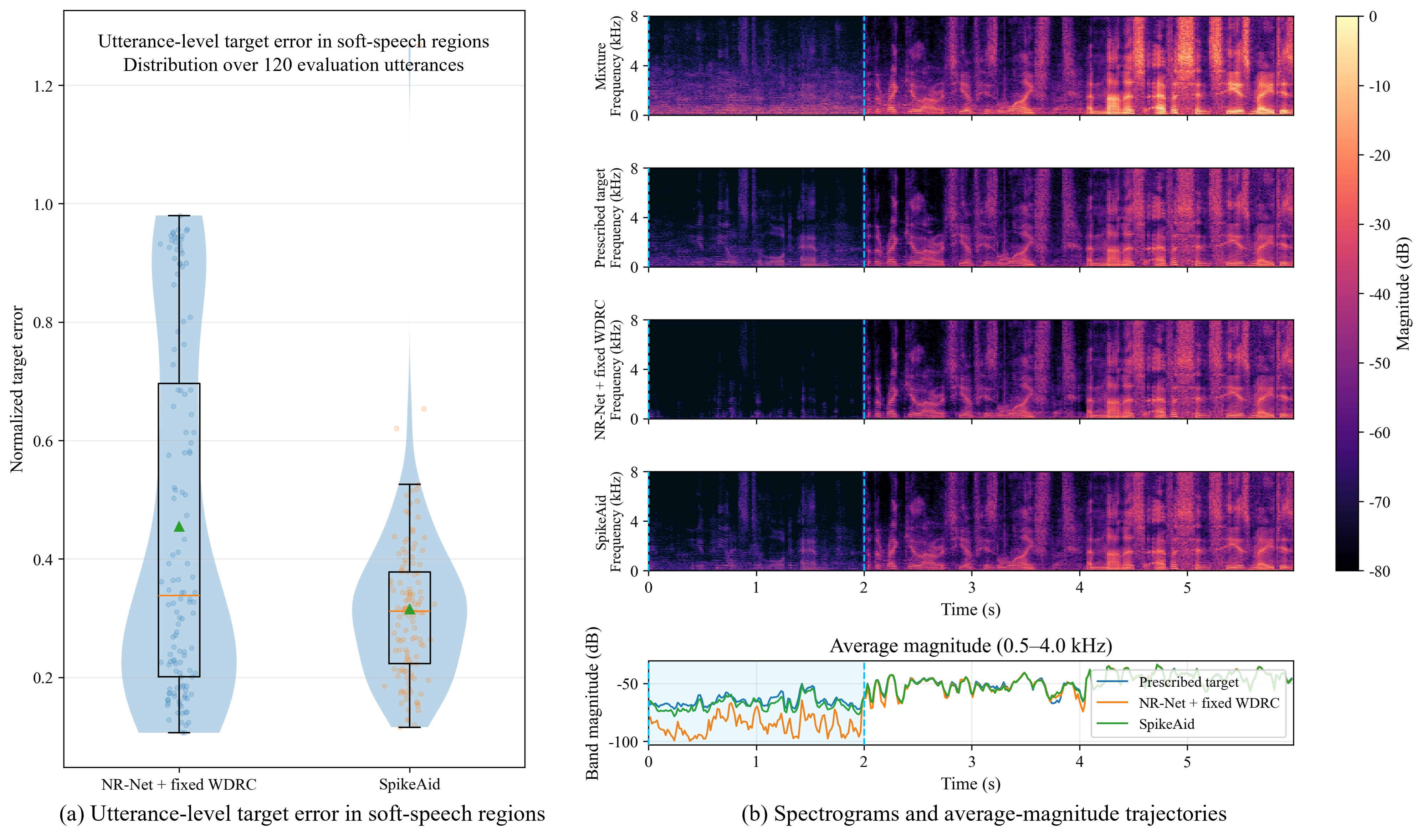}
\caption{\textbf{Target matching in soft-speech regions.}
(a) Utterance-level relative magnitude errors over the 120 evaluation utterances, computed within 0.5--4~kHz for frames whose clean-speech levels are 20--40~dB below the utterance-level maximum. Points denote utterances, boxes show interquartile ranges, horizontal lines show medians, and triangles show means.
(b) Spectrograms and 0.5--4~kHz average-magnitude trajectories for an illustrative utterance. The shaded interval marks the analyzed soft-speech region.
}
\label{fig:soft_speech_analysis}
\end{figure*}

Fig.~\ref{fig:soft_speech_analysis}(a) compares this error between ``NR-Net + fixed WDRC'' and SpikeAid over the 120 evaluation utterances. SpikeAid shows a lower error distribution and substantially fewer high-error cases, indicating more consistent preservation of the prescribed soft-speech content.

Fig.~\ref{fig:soft_speech_analysis}(b) provides an illustrative example for one utterance. During the highlighted 0--2~s interval, ``NR-Net + fixed WDRC'' strongly attenuates weak speech structures that remain present in the prescribed target. Consequently, its 0.5--4~kHz magnitude trajectory stays below the target, whereas SpikeAid follows both the target spectrogram and the corresponding magnitude trajectory more closely. In the subsequent higher-level speech region, the outputs of the two systems become more similar.

These results provide a direct example of how joint optimization improves the final hearing aid output: by considering the downstream compression during NR learning, SpikeAid is better able to preserve low-energy speech components that would otherwise be excessively suppressed by independently optimized cascaded processing.

\begin{table*}[!t]
  \centering
  \caption{
    Comparison with SNN baselines on the hearing aid joint NR-WDRC task.
  }
  \label{tab:spiking_model_comparison}

  \scriptsize
  \renewcommand{\arraystretch}{1.2}
  \setlength{\tabcolsep}{2.25pt}

  \begin{tabular}{@{}lccccccccc@{}}
    \toprule
    \multirow{3}{*}{\textbf{Architecture}} &
    \multirow{3}{*}{
      \makecell[c]{\textbf{Neuron}\\\textbf{Model}}
    } &
    \multirow{3}{*}{
      \makecell[c]{\textbf{Params}\\\textbf{(k) $\downarrow$}}
    } &
    \multicolumn{4}{c}{\textbf{Objective Performance}} &
    \multicolumn{3}{c}{\textbf{Computational Cost Proxies}} \\
    \cmidrule(lr){4-7}
    \cmidrule(lr){8-10}

    & & &
    \textbf{PESQ $\uparrow$} &
    \textbf{STOI $\uparrow$} &
    \textbf{HASPI $\uparrow$} &
    \textbf{HASQI $\uparrow$} &
    \makecell[c]{\textbf{Power Proxy}\\
                 \textbf{(MOp/s) $\downarrow$}} &
    \makecell[c]{\textbf{PDP Proxy}\\
                 \textbf{(MOp) $\downarrow$}} &
    \makecell[c]{\textbf{Energy Cost}\\
                 \textbf{($\mu$J) $\downarrow$}} \\
    \midrule

    \multicolumn{10}{@{}l}{
      \textcolor{gray}{
        \itshape Existing SNN architectures developed for temporal processing or speech enhancement.
      }
    } \\[-1pt]

    SFNN
      & LIF
      & 990
      & 1.946
      & 0.8021
      & 0.3679
      & 0.1589
      & 36.02
      & 1.15
      & 1.04 \\

    SRNN~\cite{bellec2018long}
      & LIF
      & 972
      & 2.093
      & 0.8224
      & 0.4108
      & 0.1775
      & 41.16
      & 1.32
      & 1.19 \\

    Spike-Driven Transformer~\cite{yao2024spike}
      & LIF
      & 975
      & 2.119
      & 0.8271
      & 0.3493
      & 0.1449
      & 196.61
      & 6.30
      & 5.67 \\

    Binary S4D~\cite{stan2024learning}
      & \xy{STA}
      & 960
      & 2.098
      & 0.8365
      & \underline{0.4290}
      & {0.1800}
      & 183.31
      & 5.87
      & 5.28 \\
    Spiking-FullSubNet~\cite{hao-2025-spiking-fullsubnet}
      & GSN
      & 1000
      & \underline{2.159}
      & \underline{0.8397}
      & 0.4053
      & 0.1667
      & 868.25
      & 27.81
      & 25.03 \\

    \addlinespace[2pt]
    \midrule

    \multicolumn{10}{@{}l}{
      \textcolor{gray}{
        \itshape Controlled neuron comparison under the same SpikeAid architecture.
      }
    } \\[-1pt]

    SpikeAid-LIF
      & LIF
      & 911
      & 1.989
      & 0.8071
      & 0.3827
      & 0.1643
      & {35.23}
      & {1.13}
      & {1.02} \\

    {SpikeAid-GSN}
      & GSN \cite{hao-2025-spiking-fullsubnet}
      & 962
      & 2.119
      & 0.8339
      & 0.4189
      & 0.1782
      & 59.01
      & 1.89
      & 1.70 \\

    \xy{SpikeAid-D-RF}
      & \xy{D-RF \cite{zhang2026dendritic}}
      & \xy{998}
      & \xy{1.936}
      & \xy{0.8124}
      & \xy{0.4004}
      & \xy{\underline{0.1805}}
      & \xy{34.89}
      & \xy{1.12}
      & \xy{1.01}
      \\
      
    \xy{SpikeAid-PMSN}
      & \xy{PMSN \cite{11604167}}
      & \xy{1014}
      & \xy{2.062}
      & \xy{0.8202}
      & \xy{0.4120}
      & \xy{0.1737}
      & \xy{\underline{34.09}}
      & \xy{\underline{1.09}}
      & \xy{\underline{0.98}}
      \\

    \rowcolor{gray!12}
    \textbf{SpikeAid}
      & \textbf{DRN}
      & 994
      & \textbf{2.214}
      & \textbf{0.8412}
      & \textbf{0.4365}
      & \textbf{0.1812}
      & \textbf{30.39}
      & \textbf{0.97}
      & \textbf{0.88} 
      \\
    \bottomrule
  \end{tabular}
\end{table*}

\subsection{Superior Performance and Efficiency over SNN Baselines}

Table~\ref{tab:spiking_model_comparison} compares SpikeAid with representative SNN baselines at both the architecture and spiking neuron levels.

\subsubsection{Comparison with Existing SNN Architectures}
We first benchmark the complete SpikeAid system against representative advanced SNN architectures developed for temporal processing or speech enhancement. 
\xy{These baselines include a basic spiking feedforward neural network (SFNN), a spiking recurrent neural network (SRNN) with within-layer recurrent connections \cite{bellec2018long}, the Spike-Driven Transformer \cite{yao2024spike}, Binary S4D \cite{stan2024learning} with a signed ternary activation (STA), and Spiking-FullSubNet \cite{hao-2025-spiking-fullsubnet}.}
Each baseline is reproduced and adapted to the joint NR-WDRC task using identical data-preprocessing, training, and evaluation settings.

The evaluated SNN baselines exhibit markedly different trade-offs between objective performance and computational cost. Among them, Spiking-FullSubNet achieves the strongest PESQ and STOI scores, reflecting its effectiveness as a dedicated spiking speech enhancement architecture. Binary S4D obtains the best HASPI and HASQI results among the baselines. However, these stronger baseline models incur substantially higher computational costs. In contrast, SFNN and SRNN require considerably fewer operations but achieve lower objective performance.

SpikeAid-DRN achieves the best performance across all four evaluated metrics. Compared with the strongest SNN baseline for each metric, it improves PESQ from 2.159 to 2.214 and HASPI from 0.4290 to 0.4365, while also yielding higher STOI and HASQI scores. These gains are achieved with an estimated energy cost of only 0.88~$\mu$J, which is less than one-twentieth of that required by Spiking-FullSubNet. Therefore, SpikeAid-DRN provides a more favorable balance among enhancement quality, hearing-aid-oriented perceptual performance, and computational efficiency.

\subsubsection{Effect of the Spiking Neuron Model}

\xy{We further assess the effectiveness of DRN through controlled neuron-level comparisons. Specifically, we replace DRN in SpikeAid with the LIF neuron, Gated Spiking Neuron (GSN)~\cite{hao-2025-spiking-fullsubnet}, Dendritic Resonate-and-Fire (D-RF) neuron~\cite{zhang2026dendritic}, and PMSN~\cite{11604167}, while keeping the network architecture, training procedure, and evaluation protocol unchanged. These neuron models represent different temporal processing mechanisms, including membrane integration, temporal gating, resonant dynamics, and multi-compartment dynamics.}

\xy{DRN achieves the best results across all four objective metrics. Compared with the best competing score for each metric, it improves PESQ from 2.119 to 2.214, STOI from 0.8339 to 0.8412, HASPI from 0.4189 to 0.4365, and HASQI from 0.1805 to 0.1812. It also achieves the lowest values across all three computational-cost proxies. These consistent improvements support the effectiveness of combining heterogeneous dendritic delays with resonant dynamics in DRN for capturing the temporal acoustic information essential for joint NR--WDRC processing.}


\begin{table*}[!t]
\centering
\caption{
Comparison of speech enhancement performance on the Intel N-DNS Challenge benchmark.
Superscripts $\dagger(1)$--$\dagger(4)$ indicate the official Intel N-DNS Challenge rankings.
DRN-DNS adopts the Spiking-FullSubNet backbone with GSN neurons replaced by the proposed DRN neurons, providing a controlled comparison under the same architecture.
}
\label{tab:updated_intelndns}

\resizebox{0.85\textwidth}{!}{%
\renewcommand{\arraystretch}{1.2}
\setlength{\tabcolsep}{3pt}

\begin{tabular}{lcccccccccc}
\toprule

\multirow{3}{*}{\textbf{Method}}
&
\multirow{3}{*}{
    \makecell[c]{\textbf{SI-SNR}\\
    \textbf{(dB)} $\uparrow$}
}
&
\multirow{3}{*}{
    \makecell[c]{\textbf{SI-SNRi}\\
    \textbf{(dB)} $\uparrow$}
}
&
\multicolumn{3}{c}{\textbf{DNSMOS}}
&
\multirow{3}{*}{
    \makecell[c]{\textbf{Latency}\\
    \textbf{(ms)} $\downarrow$}
}
&
\multicolumn{3}{c}{\textbf{Computational Cost Proxies}}
&
\multirow{3}{*}{
    \makecell[c]{\textbf{Params}\\
    \textbf{(k)} $\downarrow$}
}
\\

\cmidrule(lr){4-6}
\cmidrule(lr){8-10}

&
&
&
\multirow{2}{*}{\textbf{OVR} $\uparrow$}
&
\multirow{2}{*}{\textbf{SIG} $\uparrow$}
&
\multirow{2}{*}{\textbf{BAK} $\uparrow$}
&
&
\multirow{2}{*}{
    \makecell[c]{\textbf{Power Proxy}\\
    \textbf{(MOp/s)} $\downarrow$}
}
&
\multirow{2}{*}{
    \makecell[c]{\textbf{PDP Proxy}\\
    \textbf{(MOp)} $\downarrow$}
}
&
\multirow{2}{*}{
    \makecell[c]{\textbf{Energy Cost}\\
    \textbf{($\mu$J)} $\downarrow$}
}
&
\\

&
&
&
&
&
&
&
&
&
&
\\

\midrule

Noisy (Unprocessed)
& 7.37
& --
& 2.43
& 3.16
& 2.69
& --
& --
& --
& --
& --
\\

\midrule
\multicolumn{11}{l}{
\textcolor[rgb]{0.5,0.5,0.5}{
\textit{
Representative speech enhancement systems.
Results are cited from the Intel N-DNS repository~\cite{timcheck_intel_2023} and Spiking-FullSubNet~\cite{hao-2025-spiking-fullsubnet}.}
}
}
\\

Microsoft NsNet2~\cite{dubey2022icassp}
& 11.63
& 4.26
& 2.95
& 3.26
& 3.93
& \underline{20.02}
& 136.13
& 2.72
& 12.51
& 2681
\\

DCCRN~\cite{hu_dccrn_2020}
& 12.46
& 5.09
& 2.85
& 3.11
& 3.77
& \underline{20.02}
& 5070
& 100
& 460
& 1247
\\

FullSubNet~\cite{hao_fullsubnet_2021}
& 13.55
& 6.18
& 2.93
& 3.22
& 3.84
& 32.03
& 3650
& 120
& 550
& 1141
\\

Fast FullSubNet~\cite{hao_fast_2023}
& 13.88
& 6.51
& 2.91
& 3.24
& 3.82
& 32.03
& 490
& 20
& 90
& 1141
\\

CMGAN~\cite{cao_cmgan_2024}
& 14.01
& 6.64
& 2.81
& 3.23
& 3.84
& \underline{20.02}
& 15940
& 320
& 1470
& 1413
\\

\midrule
\multicolumn{11}{l}{
\textcolor[rgb]{0.5,0.5,0.5}{
\textit{
Intel N-DNS Challenge submissions.
Results and rankings are obtained from the official challenge repository~\cite{timcheck_intel_2023}.}
}
}
\\

CTDNN-LAVADL$^{\dagger(2)}$
& 13.52
& 6.59
& 2.97
& 3.32
& 3.86
& 32.00
& 61.37
& 1.96
& 1.76
& 905
\\

Sparsity-SDNN$^{\dagger(3)}$
& 12.16
& 4.80
& 2.70
& 3.19
& 3.46
& 32.01
& \textbf{9.32}
& \textbf{0.30}
& \textbf{0.27}
& {344}
\\

PSNN$^{\dagger(4)}$
& 12.32
& 4.96
& 2.68
& 2.91
& 3.96
& 32.00
& 57.24
& 1.83
& 1.65
& 724
\\

\midrule
\multicolumn{11}{l}{
\textcolor[rgb]{0.5,0.5,0.5}{
\textit{
Recent SNN-based speech enhancement systems.
Results are reported from their original papers.}
}
}
\\

Spiking-S4~\cite{chen2024spikingS4}
& 14.58
& 7.21
& 2.85
& 3.21
& 3.74
& --
& --
& --
& --
& {530}
\\

DPSNN~\cite{sun-2024-dpsnn}
& 14.70
& 7.34
& 2.90
& 3.27
& 3.77
& \textbf{10.00}
& 87.47
& \underline{0.87}
& \underline{0.78}
& 1320
\\

\midrule
\multicolumn{11}{l}{
\textcolor[rgb]{0.5,0.5,0.5}{
\textit{
Controlled comparison using the identical Spiking-FullSubNet backbone.}
}
}
\\

Spiking-FullSubNet$^{\dagger(1)}$
& \underline{15.20}
& \underline{7.83}
& \underline{3.03}
& \underline{3.35}
& \underline{3.94}
& 32.02
& 51.30
& 1.64
& 1.48
& 965
\\

\rowcolor{gray!12}
\textbf{DRN-DNS (Ours)}
& \textbf{15.73}
& \textbf{8.36}
& \textbf{3.07}
& \textbf{3.37}
& \textbf{3.98}
& 32.02
& \underline{47.49}
& 1.52
& 1.37
& 922
\\

\bottomrule
\end{tabular}
}
\end{table*}

\subsection{Enhanced Neuromorphic Speech Denoising with DRN}
\label{sec:ndns-benchmark}

The preceding experiments evaluate DRN primarily within the task-specific SpikeAid framework. To examine whether its temporal modeling capability extends beyond the proposed joint NR-WDRC formulation, we further evaluate DRN on the Intel N-DNS Challenge~\cite{timcheck_intel_2023}. The Intel N-DNS Challenge provides a public benchmark for real-time neuromorphic speech denoising, together with established ANN and SNN results and evaluation criteria covering speech quality and computational efficiency. \xy{Following the official Intel N-DNS Challenge evaluation protocol~\cite{timcheck_intel_2023}, we use the benchmark-defined metrics without modification. SI-SNR denotes the scale-invariant signal-to-noise ratio, and SI-SNRi denotes its improvement relative to the unprocessed noisy input. Deep Noise Suppression Mean Opinion Score (DNSMOS) reports three sub-metrics: OVR for overall quality, SIG for speech-signal quality, and BAK for background-noise quality.}

This evaluation serves two purposes. First, it tests whether the temporal
modeling advantage of DRN transfers from joint enhancement and compression to a conventional speech-denoising task without the WDRC target. Second, it enables comparison against recognized neuromorphic speech enhancement systems under a benchmark that is independent of our custom experimental setting. Consistent improvements on both tasks would indicate that DRN is not merely specialized to SpikeAid, but captures temporal structures that are broadly useful for streaming audio processing.

For this evaluation, we construct DRN-DNS based on the Spiking-FullSubNet backbone~\cite{hao-2025-spiking-fullsubnet}. The original GSN neurons are replaced with DRN neurons, while the remaining denoising architecture is retained. The direct comparison between DRN-DNS and Spiking-FullSubNet therefore examines the effect of the neuron model independently of the surrounding network design. Table~\ref{tab:updated_intelndns} also includes the official Intel N-DNS baselines, representative real-time ANN models, top-ranking challenge entries, and recent SNN-based speech enhancement methods to place the results within the broader benchmark context.

As shown in Table~\ref{tab:updated_intelndns}, DRN-DNS achieves the strongest overall performance among the evaluated SNN systems. In particular, compared with Spiking-FullSubNet, replacing the original GSN neurons with DRN improves SI-SNR by 0.53~dB, while increasing the DNSMOS OVR score from 3.03 to 3.07. Meanwhile, the power, PDP, and energy proxies are reduced by approximately 7.4\%. DRN-DNS also remains competitive with the ANN and challenge entries reported on the benchmark. These results demonstrate that the advantage of DRN is not confined to the SpikeAid formulation, as it improves both enhancement quality and computational efficiency when incorporated into an independent speech-denoising architecture.

\subsection{Neuromorphic Hardware Implementation and Efficiency Evaluation}

\begin{table*}[t]
    \centering
    \caption{Hardware resource utilization and power consumption of the FPGA-based SpikeAid implementation.}
    \label{tab:hardware_resource}

    \begingroup
    \renewcommand{\arraystretch}{1.15}
    \setlength{\tabcolsep}{3mm}
    \footnotesize

    \resizebox{0.85\linewidth}{!}{%
    \begin{tabular}{lcccccc}
        \toprule
        \textbf{Platform}
        & \textbf{LUTs}
        & \textbf{FFs}
        & \textbf{BRAM}
        & \textbf{DSP48}
        & \textbf{Power}
        & \textbf{Frequency} \\
        \midrule

        AXU5EV
        & \makecell{8092 / 117120  (6.90\%)}
        & \makecell{7400 / 234240  (3.16\%)}
        & \makecell{38.5 / 144  (26.74\%)}
        & \makecell{96 / 1248  (7.69\%)}
        & 2.81 W
        & 250 MHz \\

        \bottomrule
    \end{tabular}%
    }

    \endgroup
\end{table*}

\begin{table}
    \centering
    \caption{Comparison of throughput, power, and energy efficiency of SpikeAid implementations across CPU/GPU and FPGA-based neuromorphic hardware. Ratio = Neuromorphic Energy Efficiency / Corresponding Platform Energy Efficiency.
    }
    \label{tab:energy_compare}
    
    \begingroup
    \renewcommand{\arraystretch}{1.2}
    \resizebox{\linewidth}{!}{%
        \begin{tabular}{ccccc}
            \toprule
            \textbf{Device}
            & \makecell[c]{\textbf{Throughput} \\ \textbf{(Samples/s)} $\uparrow$} 
            & \makecell[c]{\textbf{Power} \\ \textbf{(W)} $\downarrow$}  
            & \makecell[c]{\textbf{Energy Efficiency} \\ \textbf{(Samples/J)} $\uparrow$}
            & \makecell[c]{\textbf{Ratio} $\downarrow$} \\
            \midrule
            
            CPU
            & 9.175
            & 19.015
            & 0.483
            & 101.9$\times$ \\
            
            GPU
            & 6.312
            & 111.166
            & 0.057
            & 865.6$\times$ \\

            \cmidrule{1-5}
            \rowcolor{gray!12}
            \textbf{Neuromorphic}
            & \textbf{138.12}
            & \textbf{2.81}
            & \textbf{49.15}
            & \textbf{1$\times$} \\
            
            \bottomrule
        \end{tabular}%
    }
    \endgroup
\end{table}

\begin{figure}[!t]
\centering
\includegraphics[width=1\linewidth]{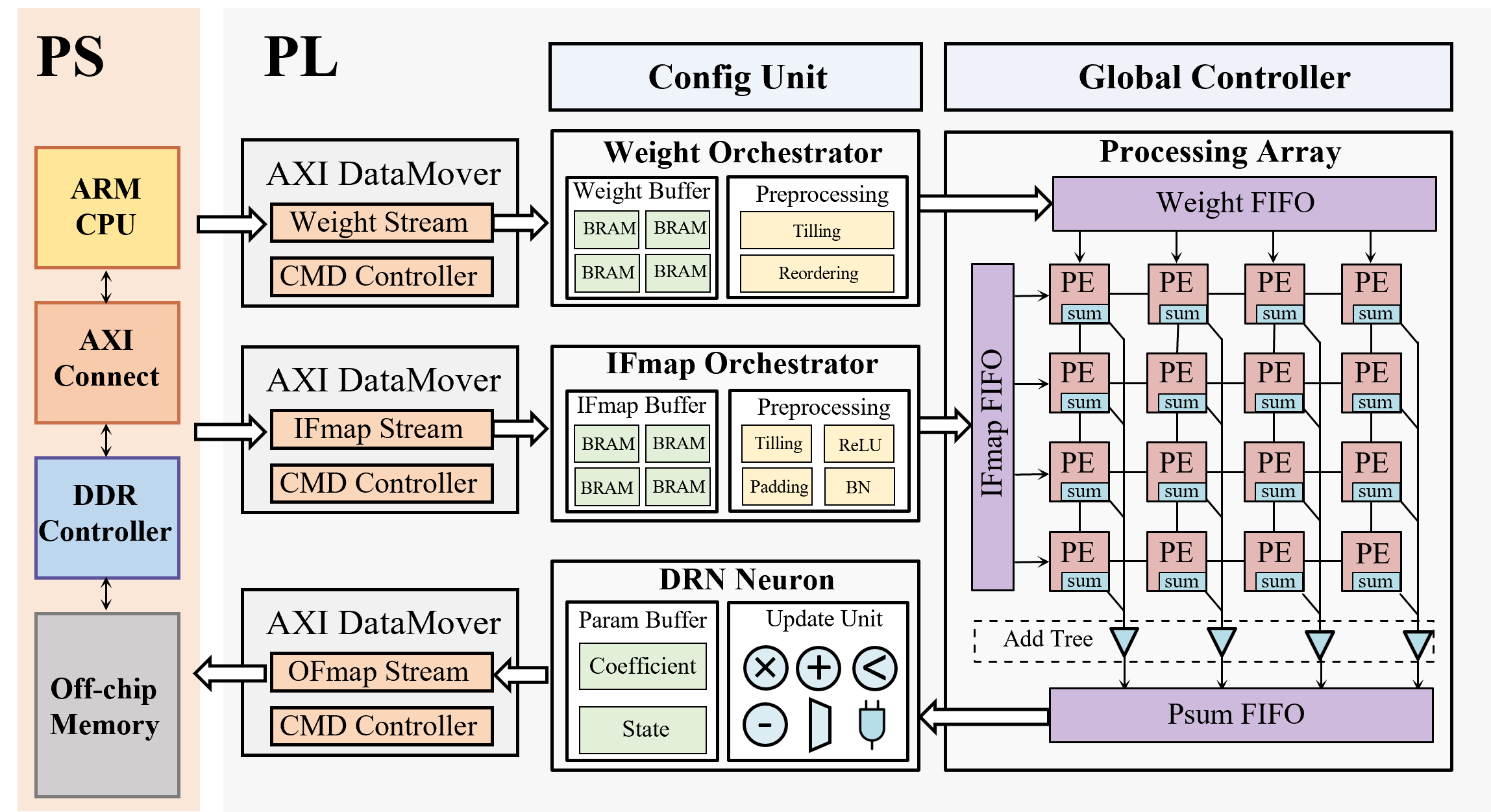}
\caption{\textbf{Overall architecture of the FPGA-based neuromorphic accelerator for SpikeAid.}
}
\label{fig:hardware_architecture}
\end{figure}

\begin{table}[t]
\centering
\small
\caption{Ablation study of the delay and resonate mechanisms in DRN. The gray-shaded row indicates the default configuration.}
\label{tab:ablation}
\renewcommand{\arraystretch}{1.4}
\resizebox{0.99\linewidth}{!}{%
    \begin{tabular}{ccccccc}
        \toprule
        \multicolumn{3}{c}{\textbf{DRN Configuration}}
        & \multicolumn{4}{c}{\textbf{Objective Performance}} \\
        \cmidrule(lr){1-3} \cmidrule(lr){4-7}
        \textbf{Resonance}
        & \textbf{Delay}
        & \textbf{$D$}
        & \textbf{PESQ $\uparrow$}
        & \textbf{STOI $\uparrow$}
        & \textbf{HASPI $\uparrow$}
        & \textbf{HASQI $\uparrow$} \\
        \hline

        \ding{55} & \ding{51} & 7 
        & 2.157 & 0.8357  & 0.4278  & 0.1799 \\

        \ding{51} & \ding{55} & -- 
        & 2.143 & 0.8364 & 0.4247 & 0.1807 \\
        \hline
        
        \ding{51} & \ding{51} & 3  
        & 2.189 & 0.8404 & 0.4273 & 0.1802 \\

        \ding{51} & \ding{51} & 5  
        & 2.196 & 0.8411 & 0.4295 & 0.1809 \\
        
        \rowcolor{gray!12}
        \ding{51} & \ding{51} & \textbf{7} 
        & \textbf{2.214} & 0.8412 & \textbf{0.4365} & 0.1812 \\

        \ding{51} & \ding{51} & 9 
        & 2.207 & \textbf{0.8432} & \underline{0.4303} & \textbf{0.1819} \\

        \ding{51} & \ding{51} & 11 
        & \underline{2.213} & \underline{0.8414} & 0.4301 & \underline{0.1818} \\

        \bottomrule
    \end{tabular}%
}

\end{table}

SpikeAid also offers practical potential for energy-efficient inference, which is critical for hearing aids that operate continuously under stringent power constraints. To evaluate its hardware efficiency, we develop a dedicated FPGA-based neuromorphic accelerator for SpikeAid, as illustrated in Fig.~\ref{fig:hardware_architecture}. The accelerator is implemented on an AXU5EV FPGA platform using the Xilinx Zynq UltraScale+ XCZU5EV device and operates at 250~MHz. The processing system (PS) handles system-level control and external-memory access, while the programmable logic (PL) performs SpikeAid inference. The PL integrates dedicated weight and input-feature orchestrators, a processing array for matrix operations, a DRN neuron unit for neuronal-state updates and spike generation, and configuration and global-control units for coordinating data transfer and layer-wise execution. Input features and weights are transferred through the Advanced eXtensible Interface (AXI) DataMover interface and processed sequentially through the computation array and DRN neuron unit, enabling causal layer-by-layer inference of SpikeAid. The accelerator is implemented at the register-transfer level using Verilog and synthesized with Xilinx Vivado 2021.1. The resulting hardware resource utilization, operating frequency, and power consumption are reported in Table~\ref{tab:hardware_resource}.

Based on this FPGA-based neuromorphic implementation, we further evaluate its streaming inference efficiency against corresponding CPU and GPU implementations in terms of throughput, power consumption, and energy efficiency, as reported in Table~\ref{tab:energy_compare}. Specifically, the CPU implementation is evaluated on an Intel Xeon Platinum 8338C processor using PyTorch 2.5.1, while the GPU implementation runs on an NVIDIA RTX 3090 with PyTorch 2.5.1 and CUDA 11.8. To reflect the operating condition of a real hearing aid device, all platforms perform causal single-stream inference by processing the input frame by frame. The average power consumption of the CPU and GPU implementations is measured during inference using platform-specific power-monitoring tools, and energy efficiency is computed as the ratio of throughput to average power consumption. 
The neuromorphic implementation achieves 101.9$\times$ and 865.6$\times$ higher energy efficiency than the CPU and GPU implementations, respectively. This improvement of approximately two orders of magnitude demonstrates the effectiveness of dedicated neuromorphic acceleration for energy-efficient streaming inference of SpikeAid.

\subsection{Ablation Study of DRN Components}

We further investigate the contributions of different components in DRN, as summarized in Table~\ref{tab:ablation}. First, either the delay mechanism or the resonate mechanism is removed from the DRN model to assess its individual contribution. Second, the number of delay branches $D$ is varied to examine how the number of temporally delayed input pathways affects local temporal integration.

\xy{Removing the delay mechanism, which reduces DRN to a resonate-and-fire neuron, degrades all four objective metrics. This comparison shows that resonance dynamics alone are insufficient for effective hearing-aid processing. Although such oscillatory dynamics support multi-timescale temporal modeling \cite{izhikevich2001resonate, pmlr-v235-higuchi24a}, without heterogeneous dendritic delays they are less effective at capturing fine-grained transient acoustic structures.
A similar degradation is observed when the resonant mechanism is removed, confirming that both components contribute to effective temporal modeling. }
Increasing $D$ from 3 to 7 leads to consistent improvements across the evaluated metrics, indicating that a richer set of delayed pathways helps the neuron capture more diverse local temporal structures. However, when $D$ is further increased to 9 or 11, the four metrics no longer exhibit consistent improvements, suggesting that the benefit of additional delay branches begins to saturate. This may be attributed to the resonating dynamics already capturing longer-range temporal dependencies, making additional delayed pathways increasingly redundant. Accordingly, $D=7$ is adopted as the default setting, providing a favorable balance between temporal modeling capability and model complexity.

\section{Conclusion}

In this work, we proposed SpikeAid, a causal end-to-end SNN framework for joint NR and WDRC in hearing aid signal processing. SpikeAid builds both NR-Net and WDRC-Net upon the proposed DRN model, which combines delayed input integration with resonating dynamics to capture temporal structures over different timescales while maintaining efficient spiking computation. The two components are further jointly optimized according to a unified hearing aid objective, allowing enhancement and compression to adapt cooperatively rather than operating as independently optimized cascaded stages.

Experimental results show that joint optimization provides clear gains over cascaded processing and improves the preservation of target-relevant speech components. Comparisons with existing SNN baselines demonstrate a favorable balance between hearing aid output quality and computational efficiency. At the neuron level, comparisons with other spiking neuron models verify the superior temporal modeling capability and efficiency of DRN. Evaluation on the public Intel N-DNS benchmark further shows that DRN achieves state-of-the-art speech-denoising performance among existing neuromorphic denoising systems with comparable computational cost.

Finally, we implement SpikeAid on FPGA-based neuromorphic hardware and evaluate its streaming inference efficiency across different hardware platforms. It achieves over 100$\times$ and 800$\times$ higher energy efficiency than the CPU and GPU implementations, respectively, demonstrating the effectiveness of dedicated neuromorphic acceleration for energy-efficient streaming inference. Future work will further integrate SpikeAid into a wearable hearing aid prototype to evaluate its real-time performance, energy efficiency, and practical robustness under realistic listening and on-device operating conditions.

\appendices

\section{Stability and Temporal-Response Analysis of DRN}

For the $j$th DRN, the continuous-time resonating dynamics can be written as 
\begin{equation} 
\frac{\mathrm{d}z_j(t)}{\mathrm{d}t} = \lambda_j z_j(t)+I_j^{d}(t), \qquad \lambda_j=\alpha_j+\mathrm{i}\beta_j. \label{eq:supp_drn_continuous_dynamic} 
\end{equation}

Under zero-order hold, the input is assumed to remain constant within each discretization interval of duration $\Delta_j$. The exact state transition is therefore 
\begin{align} z_j[t] &= \exp(\Delta_j\lambda_j)z_j[t-1] \notag\\ &\quad+ \int_{0}^{\Delta_j} \exp\!\left[ (\Delta_j-\tau)\lambda_j \right] \mathrm{d}\tau\, I_j^{d}[t] \notag\\ &= A_jz_j[t-1]+B_jI_j^{d}[t], \label{eq:supp_drn_zoh_derivation} 
\end{align} 
where 
\begin{equation} A_j = \exp(\Delta_j\lambda_j), \qquad B_j = \frac{\exp(\Delta_j\lambda_j)-1}{\lambda_j}. 
\label{eq:supp_drn_zoh_coefficients} 
\end{equation} 

The magnitude of the state-transition coefficient is 
\begin{align} 
|A_j| &= \left| \exp\!\left[ \Delta_j(\alpha_j+\mathrm{i}\beta_j) \right] \right| \notag\\ 
&= \exp(\Delta_j\alpha_j). \label{eq:supp_drn_transition_magnitude} 
\end{align} 

With the parameterization $\alpha_j=-\exp(a_j)$ and $\Delta_j>0$, we obtain 
\begin{equation} 
|A_j| = \exp\!\left[ -\Delta_j\exp(a_j) \right] <1. \label{eq:supp_drn_stability_condition} \end{equation} 

The reset-free resonating state is therefore asymptotically stable in the absence of input. Under a zero initial state, the recurrence can be unfolded as 
\begin{equation} 
z_j[t] = \sum_{i=0}^{t} A_j^{t-i}B_jI_j^{d}[i]. \label{eq:supp_drn_unfolded_recurrence} \end{equation} 

Thus, the resonating dynamics are equivalent to a causal convolution with impulse response 
\begin{equation} 
h_j^{r}[n] = B_jA_j^{n}, \qquad n\geq 0. \label{eq:supp_drn_impulse_response} 
\end{equation} 

Substituting the stable parameterization gives \begin{equation} 
h_j^{r}[n] = B_j \exp\!\left[ -n\Delta_j\exp(a_j) \right] \exp\!\left( \mathrm{i}n\Delta_j\beta_j \right),
\label{eq:supp_drn_damped_kernel} 
\end{equation} 
where the term $\Delta_j\exp(a_j)$ controls the exponential decay rate and effective memory length, whereas $\Delta_j\beta_j$ determines the phase progression of the oscillation. The corresponding discrete-frequency response is 
\begin{equation} 
H_j^{r}(e^{\mathrm{i}\omega}) = \frac{B_j} {1-A_je^{-\mathrm{i}\omega}}. \label{eq:supp_drn_frequency_response} 
\end{equation} 
Its response is concentrated around the phase of $A_j$, giving the approximate preferred temporal frequency 
\begin{equation} 
\Omega_j \approx \Delta_j\beta_j.
\label{eq:supp_drn_preferred_frequency} \end{equation} 
The DRN population with heterogeneous $a_j$, $\beta_j$, and $\Delta_j$ therefore realizes a set of damped temporal filters with different decay rates and preferred frequencies.

\section{Implementation of the Reference WDRC Operator}
\label{sec:supp_reference_wdrc}

This section specifies the deterministic WDRC operator
$\mathcal{W}_{\Theta_q}$ used to generate the source-specific reference gains. Let $q\in\{s,v\}$ index speech and noise, with
$\mathbf{Q}^{s}=\mathbf{S}$ and $\mathbf{Q}^{v}=\mathbf{V}$.

\subsection{Calibrated Band-Level Estimation}

For an $N_{\mathrm{FFT}}$-point one-sided STFT computed using analysis
window $w[n]$, the normalized power at frequency bin $f$ and frame $t$
is
\begin{equation}
    P^{q}_{f,t}
    =
    \frac{
        \kappa_f
        \left|Q^{q}_{f,t}\right|^{2}
    }{
        N_{\mathrm{FFT}}
        \sum_n w^{2}[n]
    },
    \qquad
    \kappa_f
    =
    \begin{cases}
        1, & f\in\{0,F-1\},\\
        2, & \text{otherwise}.
    \end{cases}
    \label{eq:supp_normalized_power}
\end{equation}
The factor $\kappa_f$ compensates for the omitted negative-frequency
bins. The STFT bins are assigned to $N_b$ nonoverlapping compression
bands. For a band with center frequency $f_b$, the nominal passband is
$[f_b/\sqrt{2},f_b\sqrt{2})$; the outer bands are extended to direct
current and the Nyquist frequency.

Let $\mathcal{F}_b$ denote the set of STFT bins assigned to band $b$.
The calibrated short-time band level is
\begin{equation}
    L^{q}_{b,t}
    =
    10\log_{10}
    \left[
        \max
        \left(
            \sum_{f\in\mathcal{F}_b}P^{q}_{f,t},
            \epsilon_{P}
        \right)
    \right]
    +
    L_{\mathrm{cal}},
    \label{eq:supp_band_level}
\end{equation}
where $L_{\mathrm{cal}}$ is the waveform-amplitude-to-SPL calibration
offset and $\epsilon_{P}>0$ is a numerical power floor.

\subsection{Asymmetric Envelope Tracking}

Let
$\Delta_{\mathrm{frm}}=N_{\mathrm{hop}}/f_s$
denote the frame interval. The attack and source-dependent release
coefficients are
\begin{equation}
    \gamma_{\mathrm{atk}}
    =
    \exp
    \left(
        -\frac{\Delta_{\mathrm{frm}}}
        {\tau_{\mathrm{atk}}}
    \right),
    \qquad
    \gamma_{\mathrm{rel}}^{q}
    =
    \exp
    \left(
        -\frac{\Delta_{\mathrm{frm}}}
        {\tau_{\mathrm{rel}}^{q}}
    \right).
    \label{eq:supp_compressor_coefficients}
\end{equation}

With
$\overline{L}^{q}_{b,0}=L^{q}_{b,0}$,
the causal envelope recursion is
\begin{equation}
\begin{aligned}
    \overline{L}^{q}_{b,t}
    &=
    \gamma^{q}_{b,t}
    \overline{L}^{q}_{b,t-1}
    +
    \left(
        1-\gamma^{q}_{b,t}
    \right)
    L^{q}_{b,t},
    \\
    \gamma^{q}_{b,t}
    &=
    \begin{cases}
        \gamma_{\mathrm{atk}},
        &
        L^{q}_{b,t}>
        \overline{L}^{q}_{b,t-1},
        \\
        \gamma_{\mathrm{rel}}^{q},
        &
        \text{otherwise}.
    \end{cases}
\end{aligned}
\label{eq:supp_level_smoothing}
\end{equation}

\subsection{Compression Curve and Frequency Interpolation}

A broken-stick compression curve converts the smoothed band level into
a gain in decibels:
\begin{equation}
    g^{q,\star}_{b,t}
    =
    \left(
        \frac{1}{\mathrm{CR}}-1
    \right)
    \left[
        \overline{L}^{q}_{b,t}
        -
        \mathrm{CT}
    \right]_{+},
    \qquad
    [z]_{+}=\max(z,0),
    \label{eq:supp_broken_stick_gain}
\end{equation}
where $\mathrm{CT}$ and $\mathrm{CR}$ denote the compression threshold
and compression ratio, respectively.

The band-domain gains are interpolated over the STFT frequency bins
using a natural cubic spline. The lowest and highest band gains are
extended to the direct-current and Nyquist bins before interpolation.
If $\mathcal{I}$ denotes this interpolation operator, the final linear
reference gain map is
\begin{equation}
    G^{q,\star}_{f,t}
    =
    10^{
        \mathcal{I}
        \left(
            g^{q,\star}_{b,t}
        \right)_{f}
/20
    }.
    \label{eq:supp_linear_reference_gain}
\end{equation}

All parameters used in the experiments, including the compression-band
center frequencies, attack and release times, compression threshold,
compression ratio, calibration convention, and numerical floors, are
reported in the main manuscript and the accompanying configuration
files.

\section{Computational Cost Evaluation}
\label{sec:si_computational_cost}

\subsection{Official Evaluation Framework}

We calculate the Power Proxy and power--delay-product (PDP)
Proxy according to the official algorithmic-track evaluation
methodology of the Intel Neuromorphic DNS Challenge
\cite{timcheck_intel_2023}. This methodology characterizes
computation through two neuromorphic primitives: synaptic
operations and neuron updates. Specifically,
\begin{equation}
    P_{\mathrm{proxy}}
    =
    \mathrm{SynOPS}
    +
    10\,\mathrm{NeuronOPS},
    \label{eq:si_power_proxy}
\end{equation}
where $\mathrm{SynOPS}$ and $\mathrm{NeuronOPS}$ denote the
average numbers of synaptic operations and neuron updates
per second of processed audio, respectively. The factor of
10 reflects the approximate relative energy cost of a neuron
update and a synaptic operation underlying the challenge's
Loihi-based proxy model \cite{timcheck_intel_2023}.

\subsection{Layer-Wise Firing Activity}

We measure the firing activity of each spiking layer during
inference. Let $s_{u,t,i}^{(l)} \in \{0,1\}$ denote the output
of neuron $i$ in layer $l$ at timestep $t$ for evaluation
utterance $u$. Let $N_l$ denote the number of neurons in
layer $l$, and let $T_{u,l}$ denote the number of valid
timesteps evaluated for that layer. The average firing
probability is
\begin{equation}
    r_l
    =
    \frac{
        \displaystyle
        \sum_u \sum_{t=1}^{T_{u,l}}
        \sum_{i=1}^{N_l} s_{u,t,i}^{(l)}
    }{
        \displaystyle
        N_l \sum_u T_{u,l}
    }.
    \label{eq:si_firing_probability}
\end{equation}
This quantity is expressed in spikes per neuron per timestep.
For a fixed timestep duration $\Delta t_l$, the corresponding
average firing rate in spikes per neuron per second is
\begin{equation}
    f_l = \frac{r_l}{\Delta t_l}.
    \label{eq:si_firing_rate}
\end{equation}
Only valid timesteps are included in these statistics;
padding introduced solely for batching is excluded.

\subsection{Synaptic-Operation Counting}

For a fully connected projection with $N_{\mathrm{in},p}$
inputs and $N_{\mathrm{out},p}$ outputs, this expression becomes
\begin{equation}
    C_{\mathrm{syn},p}
    =
    \left(\sum_u T_{u,p}\right)
    r_{\mathrm{in},p}
    N_{\mathrm{in},p} N_{\mathrm{out},p},
    \label{eq:si_dense_spiking_projection}
\end{equation}
where $r_{\mathrm{in},p}$ is the average presynaptic firing
probability. Each projection is counted separately, including
projections to distinct branches and recurrent projections
where applicable.

Let $T_{\mathrm{audio}}$ denote the total duration, in seconds,
of the evaluated audio. The synaptic-operation rate is
\begin{equation}
    \mathrm{SynOPS}
    =
    \frac{
        \displaystyle \sum_p C_{\mathrm{syn},p}
    }{
        T_{\mathrm{audio}}
    }.
    \label{eq:si_synops_rate}
\end{equation}

\subsection{Neuron-Update Counting}

Each neuron update at a valid timestep is counted as one
NeuronOP, irrespective of whether the neuron emits a spike.
The total neuron-update count is
\begin{equation}
    C_{\mathrm{neuron}}
    =
    \sum_u \sum_l T_{u,l} N_l,
    \label{eq:si_neuron_count}
\end{equation}
and the corresponding rate is
\begin{equation}
    \mathrm{NeuronOPS}
    =
    \frac{C_{\mathrm{neuron}}}{T_{\mathrm{audio}}}.
    \label{eq:si_neuronops_rate}
\end{equation}

Consequently, under identical neuron counts and update
schedules, the NeuronOPS contribution is unchanged across
neuron variants. Differences in the Power Proxy then arise
from activity-dependent synaptic computation. A lower proxy
does not imply a lower arithmetic cost for an individual
neuron update.

\subsection{Batch-Normalization Folding}

During deployment, batch normalization (BN) uses fixed running statistics and is folded into the preceding linear projection through standard BN folding \cite{jacob2018quantization}.
The resulting linear projection is included in the operation count. No separate BN operation is counted because its scaling and shifting have been absorbed into the projection weights and biases.

\subsection{PDP Proxy and Reference Energy Conversion}

Following the official Intel N-DNS Challenge definition
\cite{timcheck_intel_2023}, the PDP Proxy is
\begin{equation}
    \mathrm{PDP}_{\mathrm{proxy}}
    =
    P_{\mathrm{proxy}} L,
    \label{eq:si_pdp_proxy}
\end{equation}
where $L$ is the system latency in seconds. Power Proxy
and PDP Proxy are reported in MOp/s and MOp, respectively.
The latency $L$ is distinct from the timestep duration
used to determine the neuron-update rate.

In addition to these official challenge metrics, we express
the PDP Proxy as an estimated energy cost using a reference
accumulate-operation energy of $0.9$~pJ under 45-nm CMOS
technology \cite{hao-2025-spiking-fullsubnet}. We assign this
reference energy to each effective operation:
\begin{equation}
    E_{\mathrm{proxy}}
    =
    \mathrm{PDP}_{\mathrm{proxy}} e_{\mathrm{op}},
    \qquad
    e_{\mathrm{op}}=0.9~\mathrm{pJ/op}.
    \label{eq:si_energy_proxy}
\end{equation}
Equivalently,
\begin{equation}
    E_{\mathrm{proxy}}\,[\mu\mathrm{J}]
    =
    0.9\,
    \mathrm{PDP}_{\mathrm{proxy}}\,[\mathrm{MOp}].
\end{equation}

\section{Optimization and Model Selection}

All models are trained using the end-to-end loss in Eq.~(24), with equal complex- and magnitude-domain weights, i.e., $\mu=0.5$. AdamW is used with a learning rate of $10^{-3}$, weight decay of $10^{-3}$, $\beta_1=0.9$, $\beta_2=0.999$, and $\epsilon=10^{-8}$. 
{A global learning rate of $10^{-3}$ is applied to the network-level parameters, and a neuronal learning rate of $5\times10^{-5}$ is used for the DRN-specific parameters governing its internal dynamics.}
The per-device batch size is 64, and gradients are clipped to an $\ell_2$-norm of 10. {Training is performed for at most 1000 epochs. Validation is conducted every 20 epochs, and training is terminated early if the validation criterion does not improve for five consecutive validation evaluations. All models are trained on four NVIDIA RTX 3090 GPUs. Under this setup, training SpikeAid requires approximately 100 hours.}

\bibliographystyle{IEEEtran}
\bibliography{refs}

\newpage
 




\vfill

\end{document}